\documentclass[prb,aps,twocolumn,groupedaddress,nofootinbib,superscriptaddress,preprintnumbers]{revtex4-2}

\usepackage{graphicx} % Required for inserting images
\usepackage{amsthm}
\PassOptionsToPackage{dvipsnames}{xcolor}
\usepackage{tikz}
\usepackage{microtype}
\usepackage{braket}
\usetikzlibrary{patterns}
\usepackage[export]{adjustbox}
\usepackage{amsmath,amsfonts,amssymb,latexsym}
\usepackage{anyfontsize}
\usepackage{hhline}
\usepackage{bm}
\usepackage{verbatim}
\usepackage{enumitem}
\usepackage{mathrsfs}
\usepackage{slashed}
\usepackage{dsfont}
\usepackage{empheq}
\usepackage[normalem]{ulem}

\newcommand{\lan}{\langle}
\newcommand{\ran}{\rangle}
\newcommand{\ua}{\uparrow}
\newcommand{\da}{\downarrow}
\newcommand{\kb}[2]{|{#1}\rangle\langle{#2}|}

\renewcommand{\a}{\alpha}

\renewcommand{\th}{\theta}
\newcommand{\la}{\lambda}

\newcommand{\s}{\sigma}

\newcommand{\mcc}{\mathcal{C}}
\newcommand{\mck}{\mathcal{K}}

\newcommand{\mcg}{\mathcal{G}}

\newcommand{\mch}{\mathcal{H}}

\newcommand{\mcr}{\mathcal{R}}

\newcommand{\id}{\mathds{1}}

\newcommand\be            {\begin{equation}}
	\newcommand\ee            {\end{equation}}
\newcommand\ba            {\begin{aligned}}
	\newcommand\ea            {\end{aligned}}
\newcommand\bea{\begin{equation}\begin{aligned}}
		\newcommand\eea{\end{aligned}\end{equation}}

\definecolor{darkgreen}{RGB}{0,150,0}

\usepackage[colorlinks=true,linkcolor=magenta,citecolor=magenta,urlcolor=violet]{hyperref}
\renewcommand{\th}{\theta}

\usepackage{xstring}
\newcommand{\tightket}[1]{%
  \StrSubstitute{#1}{ }{ \! }[\tight@arg]%
  \ket{{\tight@arg}}%
}
\newcommand{\tightbra}[1]{%
  \StrSubstitute{#1}{ }{ \! }[\tight@arg]%
  \bra{{\tight@arg}}%
}

\makeatletter
  \newcommand{\spacedket}[1]{%
    \ket{\text{\spaced@aux#1 \relax}}%
  }
  \def\spaced@aux#1 #2\relax{%
    $#1$%
    \if\relax\detokenize{#2}\relax
    \else
      \,\spaced@aux#2\relax
    \fi
  }
\makeatother

\begin{document}

\title{Quantum fragmentation}
\author{Yiqiu Han}
\email{Yiqiu.Han@colorado.edu}
\affiliation{Department of Physics and Center for Theory of Quantum Matter, University of Colorado, Boulder, CO 80309, USA}
\author{Oliver Hart}
% \altaffiliation[Current address: ]{Quantinuum, Terrington House, 13--15 Hills Road, Cambridge CB2 1NL, UK}
\affiliation{Department of Physics and Center for Theory of Quantum Matter, University of Colorado, Boulder, CO 80309, USA}
\author{Alexey Khudorozhkov}
\affiliation{Department of Physics, Boston University, Boston, MA 02215, USA}
\author{Rahul Nandkishore}
\affiliation{Department of Physics and Center for Theory of Quantum Matter, University of Colorado, Boulder, CO 80309, USA}

\begin{abstract}
We introduce a systematic protocol for constructing quantum Hilbert-space-fragmented Hamiltonians, whose Krylov-sector structure, unlike in classically fragmented models, can be fully resolved only in an entangled basis. The protocol takes as input a classically fragmented model and uses a Rokhsar-Kivelson type construction to promote it to a quantum fragmented model. Notably, the procedure also works with non-fragmented inputs (such as Ising models). We explain how the Krylov sectors of the resulting quantum fragmented model may be labeled and counted in one dimension, and outline experimentally accessible verification of quantum fragmentation, assuming the ability to prepare specific initial states and perform tomography on reduced density matrices. We further analyze the entanglement structure of the entangled basis underlying quantum fragmentation, which sharply distinguishes it from both classical fragmentation and the trivial ``fragmentation'' of generic Hamiltonians in their eigenbasis. We also extend the construction to higher dimensions, with an explicit proof of principle example in two dimensions. We expect these results to open a new route to the systematic exploration of quantum fragmentation.
\end{abstract}

\maketitle
\tableofcontents

\section{Introduction}

Hilbert space shattering, aka Hilbert space fragmentation (HSF), is a recently discovered phenomenon~\cite{KHN, Sala2020} whereby the unitary time evolution matrix of a many-body quantum system block diagonalizes into multiple disconnected ``Krylov sectors'' within each symmetry sector, such that two states in different Krylov sectors are dynamically disconnected even if they have the same symmetry quantum numbers. HSF represents an exciting new frontier for quantum dynamics, and has drawn intense interest (see Ref.~\cite{moudgalya2022quantum} for a review). 

Most previous studies of HSF have focused on settings where the block diagonalization occurs in an unentangled (i.e., product-state) basis. This is known as ``classical fragmentation'' (CF), since analogous phenomena can arise in, e.g., purely classical Markov chain dynamics. However, in principle, HSF could also occur in an entangled basis, in which case it is known as ``quantum fragmentation'' (QF). The Temperley-Lieb model~\cite{TL1971, moudgalya2022hilbert}, the particle-conserving quantum East model~\cite{Brighi2023quantumEast} and the extended quantum breakdown model~\cite{Chen2024breakdown} provide concrete examples of quantum fragmentation in one spatial dimension. Related Hilbert-space structures have also been identified in all-to-all interacting models in cavity QED with permutation symmetry~\cite{balducci2025deephilbertspacealltoall}. However, we are not aware of a general, principled framework for the systematic construction of quantum fragmented models. As a result, quantum fragmentation remains largely unexplored and raises several natural questions. For instance, if one allows an arbitrary entangled basis, then any Hamiltonian becomes trivially “fragmented”: in its eigenbasis it decomposes into one-dimensional invariant subspaces, which is clearly not the phenomenon of interest. What criteria single out an entangled basis in which a block decomposition constitutes genuine quantum fragmentation? Is there a systematic route to constructing QF models? Is there a way to label the resulting Krylov sectors? How can QF be experimentally verified (and distinguished from CF)? Finally, can QF be meaningfully extended to systems in more than one dimension?

In this manuscript, we undertake a systematic exploration of QF and address the questions outlined above. We first introduce a principled route for constructing QF models. In particular, we show how any CF model can be ``promoted'' to a QF model, and demonstrate that this procedure can also generate QF starting from non-fragmented models, such as the transverse-field Ising model. We explain how to label the resulting Krylov sectors in one dimension through constructing an entangled basis, and propose experimentally accessible protocols to verify QF generated via our prescription. Next, we analyze the entanglement structure of this basis and find that it contains an extensive number of long-range entangled zero-energy eigenstates. At the same time, these states exhibit at most logarithmic entanglement, consistent with the non-ergodic nature of fragmented systems and in sharp contrast to the volume-law entanglement typical of eigenstates of generic Hamiltonians. This provides a clear distinction between QF, CF, and the trivial ``fragmentation'' of a Hamiltonian in its eigenbasis. Finally, we extend our discussion to two dimensions and present a concrete example of a two-dimensional QF model.

We note recent independent and complementary work by Zihan Zhou \textit{et al} ~\cite{zhou2026quantumhilbertspacefragmentation}. 
% we formalize our notion of QF by requiring that a QF model cannot be connected to any CF model by a finite-depth local unitary (FDLU) circuit

% \RN{and, at the same time, that the frozen states of the QF model have at most logarithmic entanglement, in contrast to the `volume law' entanglement of the eigenstates of a generic Hamiltonian. This crisply distinguishes QF from both CF and from the trivial `fragmentation' of a Hamiltonian in its eigenbasis}. Moreover, QF models typically host an extensive number of short-range entangled eigenstates, consistent with non-ergodic dynamics and in sharp contrast to generic Hamiltonians, which are only trivially “fragmented” into one-dimensional subspaces in their eigenbasis.

%\ollie{is conjecture the right word here? I would argue that any sensible \emph{definition} of QF should satisfy this property? Are we instead saying that we conjecture that the QF models produced by the recipe introduced herein satisfy this property?} \yiqiu{Yeah I just don't have a rigorous proof of this definition, so I'm using the word ``conjecture'', I've rewritten the argument.} 
% \ollie{Quantum fragmentation definition. Fragmentation in an entangled basis vs non-existence of connection to classically fragmented model via global, locality-preserving unitary?} \yiqiu{Added.}

\section{A systematic protocol for generating quantum fragmentation in 1D}\label{sec:protocol}
%Hilbert space fragmentation is a phenomenon related to ergodicity-breaking where the Hilbert space is fragmented into exponentially many dynamically disconnected sectors called Krylov sectors. One can further classify it into classical and quantum fragmentation. Fragmentation is classical if each Krylov sector can be spanned by a product-state basis; otherwise, it is quantum. While many classically fragmented models have been studied, the definition of quantum fragmentation remains ambiguous, with few known examples \cite{batchelor1991temperley,aufgebauer2010quantum,moudgalya2022hilbert,Chen2024breakdown}. A typical question one may ask is that since each Hamiltonian is diagonal and hence ``fragmented'' in its eigenstate basis, how do we distinguish it from quantum fragmentation?

%To better understand quantum fragmentation, we propose a generic protocol for finding the quantum fragmented version of a given model with classical fragmentation in 1D: 
\subsection{General Construction}
In this section we explain how one may systematically construct models exhibiting QF in one dimension. Our construction parallels `Rokhsar-Kivelson' models from the spin liquid literature~\cite{RK}. Specifically, consider a classically fragmented model with open boundary conditions whose Hamiltonian is of the form
\be 
H=\sum_{x=1}^{L-k} H_{x,x+k}, 
\label{eq:classical Hamiltonian}
\ee
where the local term $H_{x,x+k}$ is supported on sites $[x,x+k]$. Given a fixed product-state basis, we define the connectivity graph of $H$ as follows. Each basis state corresponds to a vertex, and two vertices $\ket{i}$ and $\ket{j}$ are connected by an edge if and only if $\lan i|H|j\ran\neq 0$. The Krylov sector is then defined as the subspace spanned by all basis states within a given connected component of this graph.

\begin{figure}[t]
    \centering
    \includegraphics[width=0.48\textwidth]{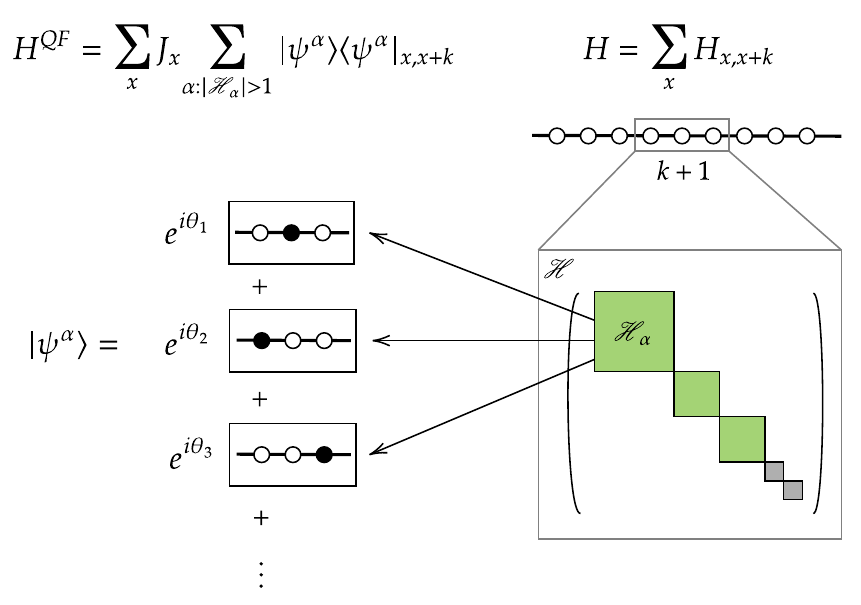}
    \caption{Illustration of the construction of QF models. Starting from a parent model shown on the right, we decompose the $(k+1)$-local Hilbert space into subspaces $\mch_\a$, each spanned by the basis $\{|w_i^\a\ran\}$. The local term in $H^{\text{QF}}$ is chosen to project onto the equal-weight (up to a phase) superposition of $\{|w_i^\a\ran\}$ for subspaces with $|\mch_\a|>1$ (highlighted in green).}
    \label{fig:QF}
\end{figure}

First, consider the limiting case where the length of the chain is $L = k+1$ and the Hamiltonian consists of a single term $H_{1,k+1}$. In this case, the Hamiltonian~\eqref{eq:classical Hamiltonian} decomposes the Hilbert space $\mch$ into disconnected subspaces \footnote{For simplicity, we ignore phenomena such as Anderson localization, where interference may prevent time evolution from exploring the full connected component.} $\mch_\alpha$, i.e.,
\be 
\mch=\bigoplus_\alpha \mch_\alpha, \quad 
\mch_\a=\text{span}\{|w_i^{\a}\ran\}.
\label{eq:classical decomp}
\ee
Here $\{\ket{w_i^\alpha}, i=1,\dots,|\mch_\a|\}$ are product states in a fixed basis, and two states $\ket{w_i^\alpha}$ and $\ket{w_j^\alpha}$ lie in the same subspace $\mathcal{H}_\alpha$ if and only if they belong to the same connected component of the connectivity graph. Using decomposition~\eqref{eq:classical decomp}, we can then construct the quantum fragmented Hamiltonian for a chain of length $L$ as
\bea H^{\text{QF}}&=\sum_{x=1}^{L-k} H^{\text{QF}}_{x,x+k} \\
&=\sum_{x=1}^{L-k}J_x\left(\sum_{\{\alpha:|\mch_\a|>1\}}|\psi^\a\ran_{x,x+k}\lan\psi^\a|_{x,x+k}\right),\label{eq: general construction}\eea
with
\be \label{eq: dimer}|\psi^\a\ran_{x,x+k} :=\sum_i e^{i\theta_i}|w_i^{\a}\ran_{x,x+k}, 
\ee
where $J_x>0$ and $\theta_i$ are arbitrary constants. The modified Hamiltonian consists of terms that project onto the (unnormalized) states $\sum_i e^{i\theta_i}|w_i^{\a}\ran_{x,x+k}$ where $\{|w_i^\a\ran_{x,x+k}\}$ are the product-state basis states between sites $x$ and $x+k$ that span the local Hilbert space $\mch_\a$ with $|\mch_\a|>1$. Note that $\theta_i$ does not have to be translation invariant, since one can always apply a depth-one local unitary circuit, amounting to a simple basis transformation, to map the Hamiltonian to an equivalent model with translation invariant $\theta_i$. It is straightforward to see that $H^{\text{QF}}$ is at least as fragmented as the parent Hamiltonian. In addition, when $L=k+1$, states of the form $e^{i\th_i}|w_i^\a\ran-e^{i\th_j}|w_j^\a\ran$ with $i\neq j$ are annihilated under $H^{\text{QF}}$, which span one-dimensional Krylov sectors. These sectors, however, cannot be spanned by a product-state basis, since the basis states $|w_i^\a\ran$ are not frozen under Hamiltonian dynamics. For larger systems, one can construct entangled frozen states according to the protocol in Sec.~\ref{sec:frozen state labeling}.  
% \AK{Is it depth-one or depth-$k$ ??}\yiqiu{Should be depth-one. Since we only need to define a local unitary on each site which assigns local phases, and make them add up to $\theta_i(x)$.} 
Consequently, whenever the parent Hamiltonian exhibits classical fragmentation, the modified model necessarily displays quantum fragmentation, justifying the superscript ``QF''. This construction, illustrated schematically in Fig.~\ref{fig:QF}, therefore provides a systematic protocol for promoting any classically fragmented model to a quantum-fragmented one.

\subsection{Models with classical fragmentation}
An example that immediately satisfies the above protocol is the Temperley-Lieb (TL) model \cite{batchelor1991temperley,aufgebauer2010quantum,moudgalya2022hilbert}, given by the Hamiltonian defined on a spin-$(N-1)/2$ chain,
\be\label{eq: TL}
H_{\text{TL}}=\sum_{x=1}^{L-1} J_x\left[\sum_{a,b=0}^{N-1}|aa\ran\lan bb|_{x,x+1}\right], \ee
which resembles the ``RK-point'' of the pair-flip (PF) model defined by,
\be H_{\text{PF}}=\sum_{x=1}^{L-1}\sum_{a,b=0}^{N-1} g^{a,b}_{x} \kb{aa}{bb}_{x,x+1},\ee
where $g_x^{a,b}$ are generic coefficients. Setting $g^{a,b}_x = J_x$, one obtains the TL model. Note that in the PF model, the local Hilbert space on two sites is fragmented into subspaces spanned by $\{|aa\ran: a=0,\dots,N-1\}$ and the frozen states $\{|ab\ran: a\neq b, a,b=0,\dots,N-1\}$. Hence, there is only one local Krylov sector with dimension greater than 1. We emphasize that $|ab\ran: a\neq b$ is only \emph{locally} frozen (i.e., it would be an eigenstate if the system had size $L=2$), but when embedded into a larger system it could acquire dynamics. 

Other classically fragmented models such as the $tJ_z$~\cite{zhang1997tjz,Batista2000tjz}, dipole-conserving~\cite{pai2019localization,khemani2020localization,sala2020ergodicity}, and particle-conserving East models~\cite{Brighi2023quantumEast,Wang2023quantumEast} involve multiple local Hilbert subspaces with dimension $|\mch_\a|>1$. Perhaps the simplest one is the $tJ_z$ model, which was originally formulated as a hard-core Fermi-Hubbard chain with only $S^z_i S^z_{i+1}$ spin interactions~\cite{zhang1997tjz,Batista2000tjz}. The onsite configuration is either empty or occupied by the spin $\s \in \{\ua,\da\}$. The allowed dynamics is
\be \ket{0\s}\leftrightarrow\ket{\s 0}, \ee
which fragments the Hilbert space into exponentially many Krylov sectors, with each sector labeled by the spin pattern. The local Hilbert space on two sites decomposes into two-dimensional subspaces spanned by $\{\ket{0\ua},\ket{\ua 0}\}$ and $\{\ket{0\da},\ket{\da 0}\}$, and other one-dimensional subspaces. As we will see in Sec.~\ref{sec:frozen state labeling}, the entangled frozen states are equal-weight superpositions of basis states within a Krylov sector, whose relative phases depend on relative distances on the connectivity graph. Note that the connectivity graph for the $tJ_z$ model with odd system sizes under PBC is non-bipartite. For example, when $L=3$, the orbit of state $|\ua00\ran$ involves $|0\ua0\ran$ and $|00\ua\ran$, whose connectivity graph is a triangle, leading to a frustration when constructing the entangled frozen states if we take the QF Hamiltonian to be a sum of local projectors onto the state $|0\s\ran_{x,x+k}+e^{i\th}|\s 0\ran_{x,x+k}$ with $\th=0$. To avoid the frustration, we set the relative phase $\th=\pi$, i.e., construct the QF Hamiltonian as
\be \label{eq: QF_tJz} H^{\text{QF}}_{tJ_z}=\sum_{x=1}^{L-1}\sum_{\a=1}^2\Pi^\a_{x,x+1},\ee
where $\Pi^\a_{x,x+1}$ are the projectors defined as
\bea \Pi^1_{x,x+1}&=(\ket{0\ua}-\ket{\ua 0})(\bra{0\ua}-\bra{\ua 0})_{x,x+1} \\
\Pi^2_{x,x+1}&=(\ket{0\da}-\ket{\da 0})(\bra{0\da}-\bra{\da 0})_{x,x+1}. \eea
States of the form $\ket{0\ua}+\ket{\ua 0}$ and $\ket{0\da}+\ket{\da 0}$ are thus frozen under the dynamics.

Finally, we consider the range-three dipole-conserving model, where the dynamics is generated by the spin-1 Hamiltonian,
	\begin{equation}
		% H_3 = \sum_j S^+_{j-1} (S_j^-)^2 S^+_{j+1} + {\rm H.c.}, \label{eq: dipole}
        H_3 = \sum_j h_j \quad \text{with} \quad h_j = S^+_{j-1} (S_j^-)^2 S^+_{j+1} + {\rm H.c.} \label{eq: dipole}
	\end{equation}
The local Hilbert space on three sites decomposes into subspaces spanned by 
% \tightket replaces space(s) with negative space
$\{\tightket{+ - +}, \tightket{0 + 0}\}$, $\{\tightket{- + -}, \tightket{0 - 0}\}$, $\{\tightket{+ - 0}, \tightket{0 + -}\}$, and $\{\tightket{- + 0}, \tightket{0 - +}\}$, 
% $\{\ket{+-+}, \ket{0+0}\}$, $\{\ket{-+-},\ket{0-0}\}$, $\{\ket{+-0},\ket{0+-}\}$, $\{\ket{-+0},\ket{0-+}\}$, 
plus locally frozen states annihilated by $h_j$.
% $S^+ (S^-)^2 S^+ + {\rm H.c.}$. 
We can transform this model into the quantum-fragmented version via Eq.~\eqref{eq: general construction},
\be \label{eq:H^QF_3} H^{\text{QF}}_3 = \sum_{x=1}^{L-2}\sum_{\a=1}^4\Pi^\a_{x,x+2}, \ee
where $\Pi^\a_{x,x+2}$ are projectors defined as
% \bea \Pi^1_{x,x+2}&=(|+-+\ran+|0+0\ran)(\lan +-+|+\lan 0+0|)_{x,x+2} \\
%      \Pi^2_{x,x+2}&=(|-+-\ran+|0-0\ran)(\lan -+-|+\lan 0-0|)_{x,x+2} \\
%      \Pi^3_{x,x+2}&=(|+-0\ran+|0+-\ran)(\lan +-0|+\lan 0+-|)_{x,x+2} \\
%      \Pi^4_{x,x+2}&=(|-+0\ran+|0-+\ran)(\lan -+0|+\lan 0-+|)_{x,x+2}. \eea
\bea \Pi^1_{x,x+2}&=(\tightket{+ - +}+\tightket{0 + 0})(\tightbra{+ - +}+\tightbra{0 + 0})_{x,x+2} \\
     \Pi^2_{x,x+2}&=(\tightket{- + -}+\tightket{0 - 0})(\tightbra{- + -}+\tightbra{0 - 0})_{x,x+2} \\
     \Pi^3_{x,x+2}&=(\tightket{+ - 0}+\tightket{0 + -})(\tightbra{+ - 0}+\tightbra{0 + -})_{x,x+2} \\
     \Pi^4_{x,x+2}&=(\tightket{- + 0}+\tightket{0 - +})(\tightbra{- + 0}+\tightbra{0 - +})_{x,x+2}. \eea
Therefore, states of the form 
% $|+-+\ran-|0+0\ran$ or $|+-0\ran-|0+-\ran$
$\tightket{+ - +}-\tightket{0 + 0}$ or $\tightket{+ - 0}-\tightket{0 + -}$
are locally frozen under the dynamics. 

For constructing entangled frozen states with larger system sizes, see Sec.~\ref{sec:frozen state labeling} and Appendix ~\ref{Appendix: other QF models}.

% \OH{general observation: there's a bit of an imbalance in the time spent on the TFIM vs the ostensibly more interesting models like the dipole-conserving model. Is there anything else interesting to say about the Hamiltonian~\eqref{eq:H^QF_3} or its eigenstates? Maybe reference the appendix?} \yiqiu{Added.}
\subsection{Transverse-field Ising model}
Interestingly, our protocol can promote even non-fragmented models into ones exhibiting QF. As an example, consider the transverse-field Ising model (TFIM) with local Hilbert space dimension $N$,
\be \label{eq: XX} H_{\text{TFIM}}=-\sum_i J_i X_i X_{i+1}-\sum_i h_i Z_i, \ee
where $J_i>0$ and $h_i$ are arbitrary constants, while $X$ and $Z$ are generalized Pauli matrices, with  $X^N=Z^N=\id$. We work in the computational $Z$-basis, where $X|n\ran=|n+1\text{ mod } N\ran$. On any bond $(i,i+1)$, the two-site Hilbert space decomposes into $N$ invariant subspaces $\mch_a$, each spanned by the orbit of $|0a\ran$ with $a=0,\dots,N-1$, under repeated applications of $X_{i} X_{i+1}$. %Since each $\mch_a$ has dimension $|\mch_a|=N>1$, this model is \emph{not} classically fragmented: the Hamiltonian connects all basis configurations within each subspace. 
Applying our protocol produces the QF Hamiltonian
\be \label{eq: QF_XX} H^{\text{QF}}_\text{TFIM} = \sum_{i=1}^{L-1}J_i\sum_{a=0}^{N-1}\Pi^a_{i,i+1} \ee
where each projector $\Pi^a_{i,i+1}=|\psi^a\ran\lan\psi^a|$ selects the equal-weight superposition over a two-site orbit
\be |\psi^a\ran=\sum_{n=0}^{N-1}|n,n+a\text{ mod } N\ran. \ee 
For example, for $N=3$,
\bea |\psi^0\ran&=|00\ran+|11\ran+|22\ran \\
     |\psi^1\ran&=|01\ran+|12\ran+|20\ran \\
     |\psi^2\ran&=|02\ran+|10\ran+|21\ran. \eea
Note that the modified Hamiltonian is positive semidefinite, so that any state belonging to the kernel of all the projectors is automatically a ground state. We show that this model is quantum fragmented by analytically determining the dimension of its zero-energy eigenspace $\mcg_L:=\ker H^{\text{QF}}_\text{TFIM}$. As proved in Appendix~\ref{App: XX},
\be \label{eq: XX_gs}|\mcg_L|=N(N-1)^{L-1}. \ee
Thus, for $N\geq 3$, the ground space exhibits exponentially large degeneracy in system size $L$. Moreover, none of these zero-energy eigenstates are product states since the local projectors annihilate no classical configurations.
%, so all ground states are genuinely entangled.
% \AK{We should probably mention somewhere that $H_{QF}$ is always positive semi-definite, and so any frozen state is a ground state. Right now, it might not be immediately clear to the reader that a zero-energy state is a ground state.}
For instance, when $N=3$ and $L=2$, the ground space is
\bea \mcg_2=\text{span}&\{|00\ran-|11\ran,|11\ran-|22\ran,|01\ran-|12\ran, \\& |12\ran-|20\ran,|02\ran-|10\ran, |10\ran-|21\ran\}, \eea
which indeed has dimension $6=3(3-1)^1$, in agreement with Eq.~\eqref{eq: XX_gs}. The exponentially many dynamically disconnected one-dimensional Krylov sectors visible only in an entangled basis strongly indicate that $H^{\text{QF}}_\text{TFIM}$ realizes genuine QF.

An intuitive way to understand the extensive ground state degeneracy is to map the Hamiltonian to a sum of single-site projectors via a basis rotation. To this end, define the ``controlled-subtraction'' unitary $U$ acting on two sites by
\be U|a,b\ran=|a,b-a\text{ mod } N\ran. \ee
When applied to $|\psi^a\ran$, this unitary yields
\be U|\psi^a\ran=\sum_n |n,a\ran:=|+\ran\otimes|a\ran, \ee
where $|+\ran:=\sum_{n=0}^{N-1}|n\ran$. Now consider the sequential circuit $\mcc:=\prod_{i=1}^{L-1}U_{i,i+1}$. Conjugating $H_{\mathrm{TFIM}}^\text{QF}$ by $\mcc$ yields
\bea \label{eq:QFXX_rotated}\mcc H^{\text{QF}}_{\text{TFIM}}\mcc^{\dagger} &= \mcc\sum_{i=1}^{L-1} J_i\sum_{a}^{N-1}|\psi^a\ran\lan\psi^a|_{i,i+1}\mcc^{\dagger} \\
&=\sum_{i=1}^{L-1} J_i|+\ran\lan+|_i. \eea
It follows immediately that the zero-energy eigenstates are precisely those that are locally orthogonal to $|+\ran$ on sites $i=1,\dots,L-1$, which reproduces the ground-state degeneracy in Eq.~\eqref{eq: XX_gs}. We emphasize that while each individual bond projector is locally unitarily equivalent to an onsite projector, implementing this rotation for all overlapping bonds requires a sequential circuit whose depth scales linearly with system size. Such circuits are capable of generating long-range correlations and long-range entanglement in quantum states~\cite{Chen2024SQC}. Consequently, $H^{\text{QF}}_{\text{TFIM}}$ cannot be related to the model in Eq.~\eqref{eq:QFXX_rotated} by any FDLU circuit. In fact, as we will see in Sec.~\ref{sec:labeling} and Sec.~\ref{sec:verify}, our construction produces long-range entangled frozen eigenstates, and so the resulting quantum fragmentation cannot be removed by any finite-depth local unitary circuit. 
% \RN{This also follows because the QF eigenstates are long range entangled (see Sec.\ref{sec:labeling}) and so cannot be related to product states by a finite depth local unitary circuit. }%\RN{How do I see that you can't do it with a finite depth brickwork circuit? Explain} \yiqiu{Maybe this claim is not right. I can only say that to connect to the CF model in Eq(19), we can't use a finite depth circuit.}

This example illustrates that quantum fragmentation need not be inherited from pre-existing classical fragmentation: even models whose dynamics is ergodic within conventional symmetry sectors (e.g., $\mathbb{Z}_N$ symmetry) can be engineered to exhibit QF via an RK-style projector construction. Moreover, there are no product states that are annihilated by the resulting QF Hamiltonian, and all frozen states are intrinsically entangled. Thus, the fragmentation in such models is purely quantum. 

\subsection{Connections to commutant algebras}
Finally, we discuss the connections of our construction of QF models to the language of bond and commutant algebras, which is a useful perspective on fragmentation~\cite{moudgalya2022hilbert}. Given a Hamiltonian which is the sum of a set of local terms $\{h_j\}$, the commutant algebra is the set of operators that commute with every term in the Hamiltonian,
\be \mcc=\{O|[O,h_j]=0,\forall j\}. \ee
In CF systems, the commutant algebra is Abelian and generated by mutually commuting integrals of motion (IOMs) which label distinct Krylov sectors. In contrast, QF systems admit irreducible representations of $\mcc$ with dimension larger than 1, which means that the commutant algebra is non-Abelian and allows degenerate Krylov subspaces. Our construction naturally generates a non-Abelian commutant algebra. For example, in the quantum fragmented $tJ_z$ model, there exist nonlocal IOMs
\be M^{\beta_1,\dots,\beta_k}_{\a_1,\dots,\a_k}=\sum_{j_1<\dots<j_k}\prod_{l=1}^k(M_{j_l})^{\beta_l}_{\a_l}, \ee
where
\be (M_j)^{\beta}_{\a}:=\ket{\beta}\bra{\a}_j, \quad \a,\beta\in\{\ua,\da\}. \ee
Such IOMs act nontrivially on the internal spin degrees of freedom while commuting with all local projectors, and are not diagonal in the product state basis if $\a_l\neq\beta_l$ for at least one $l$. Therefore, they do not mutually commute in general, implying that the commutant algebra is non-Abelian.

\section{Labeling of Krylov sectors in 1D}
\label{sec:labeling}
In this section we introduce an (over)complete labeling of Krylov sectors in QF models of the type discussed above. For convenience, we focus on models whose QF structure is layered on top of the CF structure, and we assume that each CF Krylov sector can be labeled by a classical invariant under the dynamics, e.g., using the `group word' scheme \cite{balasubramanian2023glassy}. The remaining challenge is to characterize the additional block-diagonal structure that emerges within each CF sector due to quantum fragmentation. Crucially, this finer structure is invisible in any unentangled basis, and thus cannot be captured by purely classical labeling schemes such as group words. 

\subsection{One-dimensional Krylov sectors}\label{sec:frozen state labeling}
First, we present a labeling scheme for one-dimensional Krylov sectors arising from entangled frozen states. We first consider the case where the QF Hamiltonian in Eq.~(\ref{eq: general construction}) is a sum of local projectors onto $(k+1)$-local states of the form $|w_1^\a\ran_{x,x+k}+e^{i\th}|w_2^\a\ran_{x,x+k}$, so that each non-frozen $(k+1)$-local subspace has dimension $|\mch_\a|=2$. For simplicity, we set the relative phase to $\th=0$. To construct entangled frozen states in this setting, start from a seed product state $|\phi_0^{s}\ran$ supported on the whole system and labeled by a CF label $s$.  Let $\{ \ket{\phi_a^s} \}_{a=0}^{m-1}$ be all product states dynamically connected to $\ket{\phi_0^s}$; these span the CF sector $\mck_s$. We can then construct one entangled frozen state per CF sector,
\bea 
|\psi^{s}_{\text{froz}}\ran = \sum_{a=0}^{m-1} \la_a|\phi_a^{s}\ran, \quad |\phi_a^{s}\ran \in \mck_s,
\label{eq:frozen_state}
\eea
with coefficients\footnote{For general $\theta$, $\la_a=\exp(i(\pi+\th)d_a) / \sqrt{m}$.}
\be \label{eq: lambda}\la_a = \frac{(-1)^{d_{0,a}}}{\sqrt{m}}, \ee
where $d_{0,a}$ is the distance between $|\phi_0^{s}\ran$ and $|\phi^{s}_a\ran$ on the connectivity graph (i.e., the minimal number of non-zero matrix elements of $H$ connecting them), and $m = |\mck_s|$.
Under the local projection $H^{\text{QF}}_{x,x+k}$, a basis state $|\phi_a^{s}\ran$ is mapped either to zero or to the superposition $|\phi_a^{s}\ran+|\phi_b^{s}\ran$ for a neighbor $\ket{\phi_b^s}$ on the connectivity graph with $d_{a,b}=1$, which cancels out the contribution from $H^{\text{QF}}_{x,x+k}|\phi_b^{s}\ran$ in the sum in Eq.~(\ref{eq:frozen_state}). 
Both cases lead to $H^{\text{QF}}_{x,x+k}|\psi^{s}_{\text{froz}}\ran=0$. Note that this construction is \emph{only} valid when the connectivity graph of the states within a Krylov sector is bipartite. For systems with non-bipartite connectivity graphs, the existence of cycles with an odd number of edges will lead to a frustration when constructing the entangled frozen states. This can be circumvented by constructing an alternative QF Hamiltonian with $\th=\pi$, see the previous discussion of $tJ_z$ model. For convenience, we will consider systems with bipartite connectivity graphs only. 

On the other hand, if there exists a $(k+1)$-local subspace with dimension $|\mch_\a|>2$, constructing an entangled frozen state still requires only two distinct basis states, $|w_i^\a\ran$ and $|w_j^\a\ran\in \mch_\a$ with $i\neq j$. To illustrate how this works, we use the TL model in Eq.~\eqref{eq: TL} with local Hilbert space dimension $N$ and focus on the open boundary condition (OBC). Throughout, by an ``entangled frozen state" we will mean one built from a \emph{minimal} number of basis states, so that, together with the frozen product states, it spans the kernel of the Hamiltonian.\footnote{For example, when $L=2$ and $N=2$, there is only one minimal entangled frozen state (up to an overall phase), $(|00\ran-|11\ran)/\sqrt{2}$, and other frozen states, such as $(|00\ran-|11\ran+|01\ran)/\sqrt{3}$, can be written as a linear combination of $(|00\ran-|11\ran)/\sqrt{2}$ and the frozen product state $|01\ran$.} For $L=2$ and $N\geq 3$, a generic entangled frozen state is of the form $\sum_{a=1}^N s_a|aa\ran$ with $\sum_a s_a=0$, which can be spanned by the (non-orthogonal) basis
\be \{(|a,a\ran-|a+1,a+1\ran)/\sqrt{2}, \enspace 0\leq a\leq N-2\}. \ee
Therefore, each basis state still involves only two colors.

To construct the basis of entangled frozen states for general $L$ and $N$ of the TL model under OBC, we need to first introduce the labeling scheme of the parent PF model, whose dynamics flips pairs of spins on nearest-neighboring sites with identical states. A convenient notation is to assign $N$ different colors to $N$ degrees of freedom per site, e.g., when $N=3$, we represent $|0\ran=|\textcolor{NavyBlue}{\bullet}\ran$, $|1\ran=|\textcolor{BrickRed}{\bullet}\ran$ and $|2\ran=|\textcolor{ForestGreen}{\bullet}\ran$. We connect dots with the same colors to represent dimers, e.g., $|00\ran=|\tikz[]{\draw[thick,NavyBlue] (0,0)--(0.5,0);
\fill[NavyBlue] (0,0) circle (2pt);
\fill[NavyBlue] (0.5,0) circle (2pt);}\ran$. To find the label of the Krylov sector a product state is in, one can repeat the pair-reduction procedure which removes pairs of identical spins from the bit string from left to right. The remaining unpaired spins (also known as ``dots'') form the ``irreducible string'' $s=(a_1\dots a_{L_s})$ with $a_i\neq a_{i+1}$ whose pattern is preserved under the PF dynamics and hence labels the Krylov sector of the state \cite{CahaNagaj}. For example, the state
\be
|1020110\ran=|\tikz[]{
\draw[thick,NavyBlue] (1.5,0) to[out=60, in=120] (3,0);
\draw[thick,BrickRed] (2,0)--(2.5,0);
\fill[BrickRed] (0,0) circle (2pt);
\fill[NavyBlue] (0.5,0) circle (2pt);
\fill[ForestGreen] (1,0) circle (2pt);
\fill[NavyBlue] (1.5,0) circle (2pt);
\fill[BrickRed] (2,0) circle (2pt);
\fill[BrickRed] (2.5,0) circle (2pt);
\fill[NavyBlue] (3,0) circle (2pt);
}\ran
\ee
belongs to the Krylov sector labeled by the irreducible string $s=(102)$. 

For the TL model with $L>2$ and $N\geq 3$, frozen entangled states fall into two categories: either non-separable entangled states that are annihilated by the Hamiltonian, or tensor products of frozen segments that involve distinct colors. 

We begin by constructing non-separable entangled frozen states of length $L$. As in the $L=2$ case with $N\geq3$, it suffices to restrict to basis states involving only two colors $b,c \in \{0,\dots,N-1\}$, since any state can be decomposed into linear combinations of such configurations. The labeling problem therefore reduces to the $N=2$ case. The non-separable minimal entangled frozen states can thus be labeled by an irreducible alternating string $s := (a_1 \dots a_{L_s})$, which is  either $(bcbc\dots)$ or $(cbcb\dots)$. Given such a label and a choice of colors $\{b, c\}$, these frozen states are constructed by forming a superposition
\bea \label{eq: TL frozen}|\psi^{s,\{b,c\}}_L\ran := \sum_i\la_i|\phi^{s,\{b,c\}}_{L,i}\ran, 
\eea
where the sum runs over all product states reachable from the seed state $|a_1\dots a_{L_s}b\dots b\ran$ under dynamics that involves \emph{only} the colors $\{b,c\}$. The coefficients $\lambda_i$ are the alternating signs defined in Eq.~\eqref{eq: lambda}; they enforce destructive interference among the product states, ensuring that $\ket{\psi^{s,\{b,c\}}_L}$ is annihilated by the Hamiltonian. As an example, for $L=4$ and $N=3$ we can have
\begin{equation}
    \begin{aligned}
        |\psi^{(20),\{0,2\}}_4\ran &:=\frac{1}{2}(|2000\ran-|2220\ran-|2022\ran+|0020\ran)\\
        & =: 
|\tikz[baseline={([yshift=-1.5ex]current bounding box.center)}]{
\draw[dash pattern=on 3pt off 1pt,thick,NavyBlue] (0,0.035)--(0.5,0.035);
\draw[dash pattern=on 3pt off 1pt,thick,ForestGreen] (0.0,-0.035)--(0.5,-0.035);
\draw[dash pattern=on 3pt off 1pt,thick,NavyBlue] (0.5,0.035)--(1.0,0.035);
\draw[dash pattern=on 3pt off 1pt,thick,ForestGreen] (0.5,-0.035)--(1.0,-0.035);
\draw[dash pattern=on 3pt off 1pt,thick,NavyBlue] (1,0.035)--(1.5,0.035);
\draw[dash pattern=on 3pt off 1pt,thick,ForestGreen] (1,-0.035)--(1.5,-0.035);
\fill[black] (0,0) circle (2pt);
\fill[black] (0.5,0) circle (2pt);
\fill[black] (1.0,0) circle (2pt); 
\fill[black] (1.5,0) circle (2pt); 
\node[] at (1.8,0.15) {\scriptsize $(20)$};
}\ran,
    \end{aligned}
\end{equation}
which we denote graphically by a double-dashed segment, annotated by the colors involved. Importantly, note that if the irreducible prefix spans the entire chain ($L_s = L$), the construction yields a frozen \emph{product} state. For notational convenience, we omit the superscript $\{b,c\}$ and write $|\psi^{s}_L\ran$ generically, with the understanding that: if $L_s<L$, it denotes a non-separable entangled frozen state labeled by an alternating string $s=(bcbc\dots)$ or $(cbcb\dots)$ involving only $\{b,c\}$, as in Eq.~\eqref{eq: TL frozen}; whereas if $L_s=L$, it denotes the frozen product state $|s\ran:=|a_1\dots a_{L}\ran$ with $a_i\neq a_{i+1}$.

For $N\geq 3$, one can build more elaborate frozen entangled states by taking tensor products of frozen segments that involve different colors, i.e.
\bea \label{eq: TL froz N geq 3}|\psi^{(s_1,s_2,\dots)}_{l_1,l_2,\dots}\ran:=|\psi^{s_1}_{l_1}\ran\otimes|\psi^{s_2}_{l_2}\ran\otimes\dots \eea
These compositions are constrained by color-compatibility. If both neighboring factors $|\psi^{s_{i}}_{l_i}\ran$ and $|\psi^{s_{i+1}}_{l_{i+1}}\ran$ are entangled, then they must involve disjoint sets of colors. If either of them, say $|\psi^{s_{i}}_{l_i}\ran$, is instead a product state, then its boundary color $a_{l_i}$ must not appear in the next segment $|\psi^{s_{i+1}}_{l_{i+1}}\ran$. Note that for $N=3$, only this latter type of composition --- entangled segments separated by product state segments --- guarantees that the overall state is annihilated by the TL Hamiltonian. In this case, the structure of entangled frozen states is determined not only by the labels of the PF Krylov sectors, but also by the lengths of the constituent frozen segments. 

We discuss the construction of entangled frozen states in other QF models in Appendix~\ref{Appendix: other QF models}.

\subsection{Generic Krylov sectors}
 Now, we generalize the scheme to describe the structure of generic Krylov sectors in QF models. We continue to illustrate the labeling scheme using the TL model as a concrete example. The same principles apply to generic QF models. The dynamics of the Krylov sectors can be understood in terms of a basis of dimers and frozen states \cite{moudgalya2022hilbert,Read2007Enlarged,Saito1990Proof,aufgebauer2010quantum}. A dimer between sites $i$ and $i+1$ is defined as the ``singlet'' state  
\begin{equation}
    |\psi_{\text{sing}}\ran_{i,i+1}:=\frac{1}{\sqrt{N}}\sum_{a=1}^N|aa\ran_{i,i+1}.
\end{equation}
For convenience, we use black solid lines to denote dimers, i.e., 
$|\psi_{\text{sing}}\ran=|\tikz[]{\fill[black] (0,0) circle (2pt);
\fill[black] (0.5,0) circle (2pt);
\draw[thick,black] (0,0)--(0.5,0);}
\ran.$ We thereby claim that any basis state $|\psi\ran$ can be constructed from the tensor product of $N_d$ non-crossing dimers and frozen states, i.e.,
\begin{equation}\label{eq:basis}
    |\psi\ran = |\psi_{\text{dimer}}\ran\otimes|\psi_{\text{{froz}}}\ran,
\end{equation}
where $|\psi_{\text{dimer}}\ran = \prod_{l=1}^{N_d}|\psi_{\text{sing}}\ran_{j_l,j_{l}+1}$, with $\{j_l\}$ representing the site indices of the left end of the dimers. $|\psi_{\text{froz}}\ran$ consists of entangled frozen segments annihilated by all the projectors $H_{i,i+1}^{\text{TL}}$ within the frozen regimes separated by the dimers, each of which can be labeled by the segment size and the corresponding irreducible string. 

We take $\Omega$ to be the set of basis states defined in Eq.~(\ref{eq:basis}). For example, when $L=2$ and $N=2$, 
\begin{equation}
\begin{aligned}
    \Omega(L=2)&=\left\{|\tikz[]{\fill[black] (0,0) circle (2pt);
\fill[black] (0.5,0) circle (2pt);
\draw[thick,black] (0,0)--(0.5,0);}\ran,
|\tikz[]{\fill[NavyBlue] (0,0) circle (2pt);
\fill[BrickRed] (0.5,0) circle (2pt);}\ran,
|\tikz[]{\fill[BrickRed] (0,0) circle (2pt);
\fill[NavyBlue] (0.5,0) circle (2pt);}\ran,
|\tikz[]{\draw[dash pattern=on 3pt off 1.5pt,thick,NavyBlue] (0.06,0.035)--(0.54,0.035);
\draw[dash pattern=on 3pt off 1.5pt,thick,BrickRed] (0.06,-0.035)--(0.54,-0.035);
\fill[black] (0,0) circle (2pt);
\fill[black] (0.5,0) circle (2pt);}\ran\right\}\\
    &=\left\{\frac{|00\ran+|11\ran}{\sqrt{2}},|01\ran,|10\ran,\frac{|00\ran-|11\ran}{\sqrt{2}}\right\}.
\end{aligned}
\end{equation}
In Appendix \ref{app:TL N=2}, we prove that $\Omega$ is complete and linearly independent for $N=2$. For $N\geq 3$, as mentioned in the previous section, the situation is different. Unlike in the $N=2$ case, where each frozen segment is either a product state of “dots” or a non-separable entangled frozen state, for $N\geq3$, we can also form tensor products of these building blocks. For example, $(|00\ran-|11\ran)/\sqrt{2}\otimes|2\ran$. As a result, for $N\geq3$, the set $\Omega$ becomes overcomplete: it remains complete, but is no longer linearly independent. For example, for $L=3$ and $N=3$, we have
\begin{equation}
\begin{aligned}
        &|\tikz[baseline={([yshift=-1.5ex]current bounding box.center)}]{
\draw[dash pattern=on 3pt off 1.5pt,thick,NavyBlue] (0.06,0.035)--(0.54,0.035);
\draw[dash pattern=on 3pt off 1.5pt,thick,ForestGreen] (0.06,-0.035)--(0.54,-0.035);
\draw[dash pattern=on 3pt off 1.5pt,thick,NavyBlue] (0.56,0.035)--(1,0.035);
\draw[dash pattern=on 3pt off 1.5pt,thick,ForestGreen] (0.56,-0.035)--(1,-0.035);
\fill[black] (0,0) circle (2pt);
\fill[black] (0.5,0) circle (2pt);
\fill[black] (1,0) circle (2pt); 
\node[] at (1.3,0.15) {\scriptsize (0)};
}\ran :=|000\ran-|220\ran-|022\ran \\
&=(|000\ran-|110\ran-|011\ran)+(|11\ran-|22\ran)\otimes|0\ran\\ &\quad+|0\ran\otimes(|11\ran-|22\ran)\\
&=:|\tikz[baseline={([yshift=-1.5ex]current bounding box.center)}]{
\draw[dash pattern=on 3pt off 1.5pt,thick,NavyBlue] (0.06,0.035)--(0.54,0.035);
\draw[dash pattern=on 3pt off 1.5pt,thick,BrickRed] (0.06,-0.035)--(0.54,-0.035);
\draw[dash pattern=on 3pt off 1.5pt,thick,NavyBlue] (0.56,0.035)--(1,0.035);
\draw[dash pattern=on 3pt off 1.5pt,thick,BrickRed] (0.56,-0.035)--(1,-0.035);
\fill[black] (0,0) circle (2pt);
\fill[black] (0.5,0) circle (2pt);
\fill[black] (1,0) circle (2pt);
\node[] at (1.3,0.15) {\scriptsize (0)};
}\ran+|\tikz[]{
\draw[dash pattern=on 3pt off 1.5pt,thick,BrickRed] (0.06,0.035)--(0.54,0.035);
\draw[dash pattern=on 3pt off 1.5pt,thick,ForestGreen] (0.06,-0.035)--(0.54,-0.035);
\fill[black] (0,0) circle (2pt);
\fill[black] (0.5,0) circle (2pt);
\fill[NavyBlue] (1,0) circle (2pt); 
}\ran+|\tikz[]{
\draw[dash pattern=on 3pt off 1.5pt,thick,BrickRed] (0.56,0.035)--(1.04,0.035);
\draw[dash pattern=on 3pt off 1.5pt,thick,ForestGreen] (0.56,-0.035)--(1.04,-0.035);
\fill[black] (1,0) circle (2pt);
\fill[black] (0.5,0) circle (2pt);
\fill[NavyBlue] (0,0) circle (2pt); 
}\ran.
\end{aligned}
\end{equation}
However, this does not prevent us from labeling the Krylov sectors even though the label is not unique.

To study the dynamics of the Krylov sectors, we will examine the action of the projector $\hat{P}_{i,i+1}$ on $\Omega$. By definition, it acts trivially on dimers and annihilates the state when applied within the frozen region, so we only need to consider the non-trivial action on the bonds that connect dimers and frozen segments. For convenience, we use $|\bullet\ran$ and $|\tikz[]{\draw[dash pattern=on 3pt off 1.5pt,thick,Black] (0.06,0)--(0.46,0);
\fill[black] (0,0) circle (2pt);
\fill[black] (0.5,0) circle (2pt);}\ran$ to represent dots and entangled frozen states of any color. We find that
\begin{equation}
\begin{aligned}
        \hat{P}_{1,2}|\tikz[]{\draw[thick,Black] (0.5,0)--(1,0);
\fill[black] (0,0) circle (2pt);
\fill[black] (0.5,0) circle (2pt);
\fill[black] (1,0) circle (2pt);}\ran &= |\tikz[]{\draw[thick,Black] (0.06,0)--(0.46,0);
\fill[black] (0,0) circle (2pt);
\fill[black] (0.5,0) circle (2pt);
\fill[black] (1,0) circle (2pt);}\ran \\
        \hat{P}_{2,3}|\tikz[]{
        \draw[thick,Black] (0,0)--(0.5,0);
        \draw[thick,Black] (1,0)--(1.5,0);
\fill[black] (0,0) circle (2pt);
\fill[black] (0.5,0) circle (2pt);
\fill[black] (1,0) circle (2pt);
\fill[black] (1.5,0) circle (2pt);}\ran &= |\tikz[]{
\draw[thick] (0,0) to[out=60, in=120] (1.5,0);
\draw[thick,Black] (0.5,0)--(1,0);
\fill[black] (0,0) circle (2pt);
\fill[black] (0.5,0) circle (2pt);
\fill[black] (1,0) circle (2pt);
\fill[black] (1.5,0) circle (2pt);}\ran \\
        \hat{P}_{L-3,L-2}|\dots \tikz[]{
        \draw[dash pattern=on 3pt off 1.5pt,thick,Black] (0.06,0)--(0.44,0);
        \draw[thick,Black] (1,0)--(1.5,0);
\fill[black] (0,0) circle (2pt);
\fill[black] (0.5,0) circle (2pt);
\fill[black] (1,0) circle (2pt);
\fill[black] (1.5,0) circle (2pt);}\ran &= |\dots\tikz[]{
\draw[dash pattern=on 3pt off 1.5pt,thick] (0,0) to[out=60, in=120] (1.5,0);
\draw[thick,Black] (0.5,0)--(1,0);
\fill[black] (0,0) circle (2pt);
\fill[black] (0.5,0) circle (2pt);
\fill[black] (1,0) circle (2pt);
\fill[black] (1.5,0) circle (2pt);}\ran \\
\end{aligned}
\end{equation}
To see why the above holds, note that the operator $\hat{P}_{i,i+1}$ acts nontrivially only on product states where the spins on sites $i$ and $i+1$ are identical. In particular, if there is a dimer on the bond $(i+1,i+2)$, then any constituent product state contributing a nonzero term under $\hat{P}_{i,i+1}$ must be of the form $|bbb\ran$ on sites $i$ to $i+2$,
and satisfy $\hat{P}_{i,i+1}|bbb\ran_{i,i+2}=\sum_a|aab\ran_{i,i+2}$. That is, the projector effectively swaps the $i$-th qudit with the dimer on the bond $(i+1,i+2)$. Applications of the projector can give rise to new dimers when frozen segments with colors in common are brought into adjacent positions. Consider the simplest case: a state composed of either two identical dots or two entangled frozen segments separated by a dimer. The projectors $\hat{P}_{i,i+1}$ dynamically connect such states to:
\begin{widetext}
\be
\begin{gathered}
|\tikz[]{\fill[RoyalBlue] (0,0) circle (2pt);
\fill[black] (0.5,0) circle (2pt);
\fill[black] (1,0) circle (2pt);
\fill[RoyalBlue] (1.5,0) circle (2pt);
\draw[thick,black] (0.5,0)--(1,0);}\ran \xrightarrow{\hat{P}_{1,2}}
|\tikz[]{\fill[RoyalBlue] (1,0) circle (2pt);
\fill[black] (0,0) circle (2pt);
\fill[black] (0.5,0) circle (2pt);
\fill[RoyalBlue] (1.5,0) circle (2pt);
\draw[thick,black] (0.5,0)--(0,0);}\ran \xrightarrow{\hat{P}_{3,4}}
|\tikz[]{\fill[black] (1,0) circle (2pt);
\fill[black] (0,0) circle (2pt);
\fill[black] (0.5,0) circle (2pt);
\fill[black] (1.5,0) circle (2pt);
\draw[thick,black] (0.5,0)--(0,0);
\draw[thick,black] (1,0)--(1.5,0);}\ran, \\
|\tikz[]{
\draw[dash pattern=on 3pt off 1.5pt,thick,BrickRed] (0.06,0.035)--(0.54,0.035);
\draw[dash pattern=on 3pt off 1.5pt,thick,RoyalBlue] (0.06,-0.035)--(0.54,-0.035);
\draw[thick,black] (1,0)--(1.5,0);
\draw[dash pattern=on 3pt off 1.5pt,thick,BrickRed] (2.06,0.035)--(2.54,0.035);
\draw[dash pattern=on 3pt off 1.5pt,thick,RoyalBlue] (2.06,-0.035)--(2.54,-0.035);
\fill[black] (0,0) circle (2pt);
\fill[black] (0.5,0) circle (2pt);
\fill[black] (1,0) circle (2pt); 
\fill[black] (1.5,0) circle (2pt); 
\fill[black] (2,0) circle (2pt); 
\fill[black] (2.5,0) circle (2pt);}\ran \xrightarrow{\hat{P}_{2,3}}
|\tikz[]{
\draw[dash pattern=on 3pt off 1.5pt,,thick,BrickRed] (0,0.1) to[out=60, in=120] (1.5,0.1);
\draw[dash pattern=on 3pt off 1.5pt,,thick,RoyalBlue] (0.1,0.1) to[out=50, in=130] (1.4,0.1);
\draw[thick,black] (0.5,0)--(1,0);
\draw[dash pattern=on 3pt off 1.5pt,,thick,BrickRed] (2.06,0.035)--(2.54,0.035);
\draw[dash pattern=on 3pt off 1.5pt,,thick,RoyalBlue] (2.06,-0.035)--(2.54,-0.035);
\fill[black] (0,0) circle (2pt);
\fill[black] (0.5,0) circle (2pt);
\fill[black] (1,0) circle (2pt); 
\fill[black] (1.5,0) circle (2pt); 
\fill[black] (2,0) circle (2pt); 
\fill[black] (2.5,0) circle (2pt);}\ran \xrightarrow{\hat{P}_{1,2}}
|\tikz[]{
\draw[dash pattern=on 3pt off 1.5pt,thick,BrickRed] (1.06,0.035)--(1.54,0.035);
\draw[dash pattern=on 3pt off 1.5pt,thick,RoyalBlue] (1.06,-0.035)--(1.54,-0.035);
\draw[thick,black] (0,0)--(0.5,0);
\draw[dash pattern=on 3pt off 1.5pt,thick,BrickRed] (2.06,0.035)--(2.54,0.035);
\draw[dash pattern=on 3pt off 1.5pt,thick,RoyalBlue] (2.06,-0.035)--(2.54,-0.035);
\fill[black] (0,0) circle (2pt);
\fill[black] (0.5,0) circle (2pt);
\fill[black] (1,0) circle (2pt); 
\fill[black] (1.5,0) circle (2pt); 
\fill[black] (2,0) circle (2pt); 
\fill[black] (2.5,0) circle (2pt);}\ran \\
\xrightarrow{\hat{P}_{4,5}}
\big(|\tikz[]{
\draw[thick,Black] (0,0)--(0.5,0);
\draw[thick,Black] (1.5,0)--(2,0);
\fill[black] (0,0) circle (2pt);
\fill[black] (0.5,0) circle (2pt);
\fill[BrickRed] (1,0) circle (2pt);
\fill[black] (1.5,0) circle (2pt);
\fill[black] (2,0) circle (2pt); 
\fill[BrickRed] (2.5,0) circle (2pt);}\ran +
|\tikz[]{
\draw[thick,Black] (0,0)--(0.5,0);
\draw[thick,Black] (1.5,0)--(2,0);
\fill[black] (0,0) circle (2pt);
\fill[black] (0.5,0) circle (2pt);
\fill[RoyalBlue] (1,0) circle (2pt);
\fill[black] (1.5,0) circle (2pt);
\fill[black] (2,0) circle (2pt); 
\fill[RoyalBlue] (2.5,0) circle (2pt);}\ran\big)/\sqrt{2} \\
\xrightarrow{\hat{P}_{3,4}}
|\tikz[]{
\draw[thick,Black] (0,0)--(0.5,0);
\draw[thick,Black] (1,0)--(1.5,0);
\fill[black] (0,0) circle (2pt);
\fill[black] (0.5,0) circle (2pt);
\fill[black] (1,0) circle (2pt);
\fill[black] (1.5,0) circle (2pt);}\ran\otimes
\big(|\tikz[]{
\fill[BrickRed] (0,0) circle (2pt);
\fill[BrickRed] (0.5,0) circle (2pt);}\ran+
|\tikz[]{
\fill[RoyalBlue] (0,0) circle (2pt);
\fill[RoyalBlue] (0.5,0) circle (2pt);}\ran\big)/\sqrt{2}\xrightarrow{\hat{P}_{5,6}}
|\tikz[]{
\draw[thick,Black] (0,0)--(0.5,0);
\draw[thick,Black] (1,0)--(1.5,0);
\draw[thick,Black] (2,0)--(2.5,0);
\fill[black] (0,0) circle (2pt);
\fill[black] (0.5,0) circle (2pt);
\fill[black] (1,0) circle (2pt);
\fill[black] (1.5,0) circle (2pt);
\fill[black] (2,0) circle (2pt); 
\fill[black] (2.5,0) circle (2pt);}\ran 
\end{gathered}
\ee
\end{widetext}
Subsequent applications of the projector will no longer change the total number of dimers but may rearrange their positions. Therefore, one convenient way to label a Krylov sector is to use the label of the state with the largest number of dimers within that sector.\footnote{Note that this does not violate the Hermiticity of the Hamiltonian, as the entangled basis we adopt is non-orthogonal. For example, even though applying $H$ to state $|\tikz[]{\fill[black] (1,0) circle (2pt);
\fill[black] (0,0) circle (2pt);
\fill[black] (0.5,0) circle (2pt);
\fill[black] (1.5,0) circle (2pt);
\draw[thick,black] (0.5,0)--(0,0);
\draw[thick,black] (1,0)--(1.5,0);}\ran$ can only rearrange dimers, this state is still dynamically connected to the state $|\tikz[]{
\draw[dash pattern=on 3pt off 1.5pt,thick,BrickRed] (0.06,0.035)--(0.54,0.035);
\draw[dash pattern=on 3pt off 1.5pt,thick,RoyalBlue] (0.06,-0.035)--(0.54,-0.035);
\draw[dash pattern=on 3pt off 1.5pt,thick,BrickRed] (1.06,0.035)--(1.54,0.035);
\draw[dash pattern=on 3pt off 1.5pt,thick,RoyalBlue] (1.06,-0.035)--(1.54,-0.035);
\fill[black] (0,0) circle (2pt);
\fill[black] (0.5,0) circle (2pt);
\fill[black] (1,0) circle (2pt); 
\fill[black] (1.5,0) circle (2pt); 
}\ran$ because $\lan\tikz[]{
\draw[thick] (0,0) to[out=60, in=120] (1.5,0);
\draw[thick,black] (0.5,0)--(1,0);
\fill[black] (0,0) circle (2pt);
\fill[black] (0.5,0) circle (2pt);
\fill[black] (1,0) circle (2pt);
\fill[black] (1.5,0) circle (2pt);}|\tikz[]{
\draw[dash pattern=on 3pt off 1.5pt,thick,BrickRed] (0.06,0.035)--(0.54,0.035);
\draw[dash pattern=on 3pt off 1.5pt,thick,RoyalBlue] (0.06,-0.035)--(0.54,-0.035);
\draw[dash pattern=on 3pt off 1.5pt,thick,BrickRed] (1.06,0.035)--(1.54,0.035);
\draw[dash pattern=on 3pt off 1.5pt,thick,RoyalBlue] (1.06,-0.035)--(1.54,-0.035);
\fill[black] (0,0) circle (2pt);
\fill[black] (0.5,0) circle (2pt);
\fill[black] (1,0) circle (2pt); 
\fill[black] (1.5,0) circle (2pt); 
}\ran\neq 0$.}
Below, we provide an algorithmic way to determine such a representative state for an arbitrary Krylov sector. 
% \AK{I find the rest of this subsection quite heavy on technical details, which is ok, but halfway through I got lost in what exactly we are trying to achieve. So, I think here we should clearly state which goal the remainder of the subsection achieves. The last sentence in this paragraph is my attempt at doing this. Can you verify that this sentence is correct? Is this indeed what you are doing below?}\yiqiu{Thanks for pointing it out. Looks good.}

Consider the dynamics of a state which is the tensor product of two entangled frozen segments $|\psi^{s}_{l}\ran\otimes|\psi^{s'}_{l'}\ran$ where the irreducible strings $s=(a_1\dots a_{L_s})$ with $a_i\in\{b,c\}$, and $s'=(a_1'\dots a'_{L_{s'}})$ with $a_i'\in\{d,e\}$. Assume that the color sets intersect, i.e. $\{b,c\}\cap\{d,e\}\neq\emptyset$ (otherwise, the state is simply frozen under the dynamics). In general, such a state is always connected to configurations with the frozen regime partially labeled by the irreducible label $\text{irr}(ss')$ of the concatenated string $(ss') := (a_1\dots a_La'_1\dots a'_L)$. The detailed outcome depends on how their color sets and irreducible labels overlap; assuming without loss of generality that $l>l'$, we classify the resulting dynamics as follows:

\begin{enumerate}
    \item \textbf{Complete fusion (identical color sets):}\\ 
    If $\{b,c\}=\{d,e\}$,  the state is connected to 
    \begin{enumerate}
        \item $|\psi^{\text{irr}(ss')}_{l-l'}\ran\otimes(|\tikz[]{
\draw[thick,Black] (0,0)--(0.5,0);
\fill[black] (0,0) circle (2pt);
\fill[black] (0.5,0) circle (2pt);}\ran)^{\otimes l'}$, if $|\text{irr}(ss')|< l-l'$;
        \item $|\text{irr}(ss')\ran\otimes(|\tikz[]{
\draw[thick,Black] (0,0)--(0.5,0);
\fill[black] (0,0) circle (2pt);
\fill[black] (0.5,0) circle (2pt);}\ran)^{\otimes \frac{l+l'-|\text{irr}(ss')|}{2}}$, otherwise.
    \end{enumerate}
    Note that $l+l'-|\text{irr}(ss')|$ is always even because irr$(s)$ always has the same parity as $l$.
    \item \textbf{Fusion blocking (disjoint leading color):}\\
    If $a_1'\notin\{b,c\}$, the state is connected to
    \begin{enumerate}
        \item $|\psi^{s}_{l-l'+L_{s'}}\ran\otimes|s'\ran\otimes(|\tikz[]{
\draw[thick,Black] (0,0)--(0.5,0);
\fill[black] (0,0) circle (2pt);
\fill[black] (0.5,0) circle (2pt);}\ran)^{\otimes l'-L_{s'}}$, if $L_s-L_{s'}<l-l'$;
    \item $|s\ran\otimes|\psi^{s'}_{l'-l+L_s}\ran\otimes(|\tikz[]{
\draw[thick,Black] (0,0)--(0.5,0);
\fill[black] (0,0) circle (2pt);
\fill[black] (0.5,0) circle (2pt);}\ran)^{\otimes l-L_{s}}$, otherwise.
    \end{enumerate}
    \item \textbf{Partial fusion (single-color overlap):}\\ 
    If $a_1'\in\{b,c\}$ but $a_2'\notin\{b,c\}$, the state is connected to
    \begin{enumerate}
    \item $|\psi^{\text{irr}(s a'_1)}_{l-l'+L_{s'}-1}\ran\otimes|a'_2\dots a'_{L_{s'}}\ran\otimes(|\tikz[]{
\draw[thick,Black] (0,0)--(0.5,0);
\fill[black] (0,0) circle (2pt);
\fill[black] (0.5,0) circle (2pt);}\ran)^{\otimes l'-L_{s'}+1}$, if $|\text{irr}(s a'_1)|<l-l'+L_{s'}$;
\item $|a_1\dots a_{L_s-1}\ran\otimes|\psi^{\text{irr}(a_{L_s}s')}_{l'-l+L_s-1}\ran\otimes(|\tikz[]{\fill[black] (0,0) circle (2pt);
\fill[black] (0.5,0) circle (2pt);
\draw[thick,black] (0,0)--(0.5,0);}\ran)^{\otimes l-L_{s}+1}$, otherwise.
    \end{enumerate}
\end{enumerate}
A detailed proof is provided in Appendix \ref{app: dynamics of two frozen segments}. 

For example, the state $|\psi^{(10)}_{4}\ran\otimes|\psi^{(01)}_{2}\ran$ is the tensor product of two frozen segments with the same colors $\{0,1\}$. Under the dynamics,
\begin{equation}
    \begin{aligned}
        &|\psi^{(10)}_{4}\ran\otimes|\psi^{(01)}_{2}\ran\sim\\
        &(|1000\ran-|1110\ran-|1011\ran+|0010\ran)\otimes|01\ran\\
        \to&(|100\ran-|111\ran+|001\ran)\otimes|1\ran\otimes|\tikz[]{
\draw[thick,Black] (0,0)--(0.5,0);
\fill[black] (0,0) circle (2pt);
\fill[black] (0.5,0) circle (2pt);}\ran\\
\to &(|00\ran-|11\ran)\otimes(|\tikz[]{
\draw[thick,Black] (0,0)--(0.5,0);
\fill[black] (0,0) circle (2pt);
\fill[black] (0.5,0) circle (2pt);}\ran)^{\otimes 2}\sim |\psi^{(\emptyset)}_{2}\ran\otimes(|\tikz[]{
\draw[thick,Black] (0,0)--(0.5,0);
\fill[black] (0,0) circle (2pt);
\fill[black] (0.5,0) circle (2pt);}\ran)^{\otimes 2},
    \end{aligned}
\end{equation}
is mapped to the tensor product of an entangled frozen state with length $l-l'=2$ labeled by the string irr$(1001)=(\emptyset)$ and two dimers, which verifies the argument for Case 1(a) above. If we replace the second segment by $|\psi_2^{(02)}\ran=|02\ran$, the state is instead mapped to
\bea &(|100\ran-|111\ran+|001\ran)\otimes|2\ran\otimes|\tikz[]{
\draw[thick,Black] (0,0)--(0.5,0);
\fill[black] (0,0) circle (2pt);
\fill[black] (0.5,0) circle (2pt);}\ran \\
&\sim |\psi^{(1)}_3\ran\otimes|2\ran \otimes|\tikz[]{
\draw[thick,Black] (0,0)--(0.5,0);
\fill[black] (0,0) circle (2pt);
\fill[black] (0.5,0) circle (2pt);}\ran, \eea
which is a tensor product of an entangled frozen state with length $l-l'+L_{s'}-1=3$ labeled by the string irr$(100)=(1)$, a product state $|2\ran$, and a dimer. This illustrates partial fusion, where only the leading color in $s'$ participates in the irreducible string reduction process.

To conclude, a Krylov sector of the TL model is labeled by a state of the form
\begin{equation}
|\psi^{s_1,s_2,\dots}_{l_1,l_2,\dots}\ran\otimes(|\tikz[]{
\draw[thick,Black] (0,0)--(0.5,0);
\fill[black] (0,0) circle (2pt);
\fill[black] (0.5,0) circle (2pt);}\ran)^{\otimes k},
\end{equation}
where if two adjacent frozen segments $|\psi^{s_i}_{l_i}\ran$ and $|\psi^{s_{i+1}}_{l_{i+1}}\ran$ are entangled, their color sets must be disjoint. If either segment is a product state, say the right segment, then its first spin must lie outside the color set of the left one. 

We conjecture that for a generic QF model with the Hamiltonian given in Eq.~(\ref{eq: general construction}), its Krylov sectors can be labeled by similar representation in terms of tensor products of dimers and frozen segments. In particular, when the model contains multiple local Hilbert subspaces $\mch_\a$ with $|\mch_\a|>1$, distinct species of dimers appear, labeled by $\a$ as in Eq.~(\ref{eq: dimer}). 

\subsection{Counting Krylov sectors}
With the labelling scheme constructed above, we are able to determine the total number of Krylov sectors of QF models by counting the number of states that are tensor products of dimers and a frozen state. We continue to take the TL model as an example. It is known that for the TL model on an open chain of size $L$, the zero-energy eigenspace $\mcg_L:=\ker H_{\text{TL}}$ has dimension \cite{han2026HSF}
\be\label{eq:TL ker dim} |\mcg_L| =
			\begin{cases} 
           L+1, N=2\\
           \frac{(N+\sqrt{N^2-4})^{L+1} - (N-\sqrt{N^2-4})^{L+1}}{2^{L+1} \sqrt{N^2-4}}, N>2
            \end{cases} \ee
which is equal to the number of linearly independent frozen states. 

We first consider the case when $N=2$. For a configuration with $N_d$ dimers and a frozen segment of length $L-2N_d$, there are 2 frozen product states, and $L-2N_d-1$ entangled frozen states which are non-separable, giving rise to a total of $L-2N_d+1$ possible frozen states, consistent with the result above. The total number of Krylov sectors for $N=2$ is thus
\be \#\mck_{\text{TL}}=\sum_{N_d=0}^{\lfloor\frac{L}{2}\rfloor}(L-2N_d+1)=
    \frac{(L+2)^2-\sigma}{4},
\ee
where $\sigma=L\text{ mod }2$. In comparison, the PF model with $N=2$ contains only $L+1$ Krylov sectors, each spanned by a product state basis. 

For $N>2$, the total number of Krylov sectors is
\begin{widetext}
    \bea \#\mck_{\text{TL}}&=\sum_{N_d=0}^{\lfloor\frac{L}{2}\rfloor}\frac{(N+\sqrt{N^2-4})^{L-2N_d+1} - (N-\sqrt{N^2-4})^{L-2N_d+1}}{2^{L-2N_d+1} \sqrt{N^2-4}}\\
&= \frac{q^{L+3}+q^{-L-1}-q^{1+\sigma}-q^{1-\sigma}}{(q^2-1)(2q-N)}, \eea
where
\be q=\frac{N+\sqrt{N^2-4}}{2}. \ee
\end{widetext}
In comparison, we can count the total number of Krylov sectors of the PF model with $N>2$ by counting the number of irreducible strings
\bea \#\mck_{\text{PF}}&=\begin{cases}
    \sum_{k=0}^{\frac{L-1}{2}}N(N-1)^{2k}, L \text{ odd}\\
    \sum_{k=1}^{\frac{L}{2}}N(N-1)^{2k-1}+1, L \text{ even}
\end{cases}\\
&=\frac{(N-1)^{L+1}-1}{N-2}.\eea 
The ratio of the number of Krylov sectors in the TL model to that in the PF model is thus,
\be \frac{\#\mck_{\text{TL}}}{\#\mck_{\text{PF}}}\sim\left(\frac{N+\sqrt{N^2-4}}{2N-2}\right)^{L+1}, \ee 
which grows exponentially in system size since $N+\sqrt{N^2-4}>2N-2$ for $N>2$. This reflects the enhanced fragmentation structure in the TL model, whose degeneracy in the energy spectrum is exponentially larger than that of the PF model.

\section{Verifying quantum fragmentation}\label{sec:verify}
Our construction of generic QF models allows us to use entangled frozen states to verify quantum fragmentation. Given a QF model whose Hamiltonian satisfies Eq.~(\ref{eq: general construction}), the $(k+1)$-site reduced density matrix of a minimal \emph{non-separable} entangled frozen state $|\psi^{s}_{\text{froz}}\ran$ on any interval $[x,x+k]$ takes the form
\bea \rho_{[x,x+k]}&:=\text{Tr}_{[1,x-1]\cup[x+k+1,L]}(|\psi^{s}_{\text{froz}}\ran\lan\psi^{s}_{\text{froz}}|)\\
&=\frac{c_0}{2}(e^{i\th_i}|w_i^\a\ran-e^{i\th_j}|w_j^\a\ran)(e^{-i\th_i}\lan w_i^\a|-e^{-i\th_j}\lan w_j^\a|)\\
&\quad +\sum_{n=1} c_n|\phi_n\ran\lan\phi_n|, \eea
where the first term with $i\neq j$ corresponds to an entangled frozen state supported on $k+1$ sites, and the remaining terms are superpositions of frozen product states over $k+1$ sites, with detailed structure determined by the position $x$ and the label $s$ of the state. The coefficients satisfy the normalization condition $\sum_{n=0} c_n = 1$. This construction guarantees that $H^{\text{QF}}_{x,x+k} |\psi^{s}_{\text{froz}}\rangle = 0$ for any $x\in[1,L-k]$, and that $|\psi^{s}_{\text{froz}}\rangle$ is non-separable. The nonzero overlap between an entangled (frozen) state and the reduced density matrix over \emph{any} $(k+1)$-site subsystem  reveals a distinctive feature of our construction of quantum fragmentation. Note that this method does require the ability to prepare the system in the minimal non-separable entangled frozen state, and also to do tomography on a $(k+1)$-site reduced density matrix. 

We illustrate this procedure using the TL model as an example. An especially interesting case arises when $N=2$, $L$ is \emph{odd}, and the system is subject to periodic boundary conditions (PBC). In this setting, there are only two irreducible strings, $s=(0)$ and $s=(1)$. This is because any configuration with more than one `1' -- for instance, $|10100\ran$ -- always contains a pair of `1's separated by an even number of `0's under PBC, which can be removed through the pair-reduction procedure. Consequently, there exist only two minimal entangled frozen states labeled by these two irreducible strings. One of them is given by

\bea |\psi_+\ran &= \bigotimes_{i=1}^L S_i |\text{GHZ}^{(x)}_+\ran \\
                            &:=\frac{1}{\sqrt{2}}\bigotimes_{i=1}^L S_i(|++\dots+\ran+|--\dots-\ran),
\eea
where $|\pm\ran=(|0\ran\pm|1\ran)/\sqrt{2}$, $|\text{GHZ}^{(x)}_+\ran$ is the GHZ state in the $x$ basis, $S_i$ is the S gate acting on site $i$, assigning a $\pi/2$ phase to $|1\ran$. Therefore, $|\psi_+\ran$ is a superposition of all configurations with an even number of `1's, where configurations containing $4n+2$ `1's (for $n\in\mathbb{N}$) acquire a phase of $\pi$. The other entangled frozen state, with odd parity, is given by
\bea |\psi_-\ran &= \bigotimes_{i=1}^L X_i|\psi_+\ran= \bigotimes_{i=1}^L S_i |\text{GHZ}^{(x)}_-\ran \\
                            &:=\frac{1}{\sqrt{2}}\bigotimes_{i=1}^L S_i(|++\dots+\ran-|--\dots-\ran).
\eea
Interestingly, the symmetric and antisymmetric combinations $(|\psi_+\ran\pm|\psi_-\ran)/\sqrt{2}$ are also frozen, but they are product states in the $y$ basis. 

Furthermore, their reduced density matrices of any two neighboring sites $(j,j+1)$ are always of the same form, 
\begin{equation}
\begin{aligned}
    \rho_{j,j+1}&=\text{Tr}_{i\neq j, j+1}(|\psi_+\ran\lan\psi_+|)\\
    &=\frac{1}{2}S_1S_2[|++\ran\lan++|+|--\ran\lan--|]S_1^\dagger S_2^\dagger\\
    &=\frac{1}{2}(|\phi_0\ran\lan\phi_0|+|\phi_1\ran\lan\phi_1|),
\end{aligned}
\end{equation} 
where
\bea |\phi_0\ran &= \frac{1}{\sqrt{2}}(|00\ran-|11\ran),\\
|\phi_1\ran &= \frac{1}{\sqrt{2}}(|01\ran+|10\ran) \eea
are two Bell states, with $|\phi_0\ran$ the entangled frozen state supported on 2 sites and $|\phi_1\ran$ the superposition of frozen product states. We can then use the `Bellness' of $\rho_{j,j+1}$ to verify the quantum fragmentation. 
% \ollie{what does it mean practically to use the Bellness as a diagnostic, especially if it's a statistical mixture of two different ones?}
% \yiqiu{I think the second Bell state is not important because it changes over different entangled frozen states and RDMs of different intervals $[x,x+k]$. But $|\phi_0\ran\lan\phi_0|$ is always a $k+1$ site entangled frozen state. Also $\lan\phi_0|\phi_1\ran=0$.} 

\section{Entanglement structure of quantum fragmentation}
\label{Sec: entanglement}

%\RN{I think this appendix should be `promoted' to main text. Combine it with your comment currently at the end of Section IV, and make it its own section. Maybe call it something like `entanglement structure of quantum fragmentation' and tie back to the question (raised in introduction) of how do we distinguish QF from the trivial fragmentation of a Hamiltonian in its eigenbasis (which the results of this appendix do very cleanly).}
In this section, we study the entanglement structure of the entangled basis used in our construction, which (we show) can be used to crisply distinguish QF from both CF and the trivial fragmentation of a Hamiltonian in its eigenbasis. We build on the discussion from Section \ref{sec:frozen state labeling}. Since these states are composed of tensor products of dimers and frozen segments as shown in Eq.~\eqref{eq:basis}, their bipartite entanglement entropy is strongly constrained.

When the bipartition cuts between independent segments or across dimers, the entanglement entropy is $O(1)$. The case that can produce the largest entanglement occurs when there is a non-separable long-range entangled frozen state $|\psi^s_L\ran$. Such a frozen state is an equal-weight superposition (up to some phase) of all product states in the CF Krylov sector labeled by $s$, as defined in Eq.~\eqref{eq:frozen_state}. Without loss of generality, we assume that $L$ is even. To bound the entanglement entropy, we consider a bipartition of the chain into two halves $A=[1,L/2]$ and $B=[L/2+1,L]$. Since the Schmidt rank is symmetric under $|s|\to L-|s|$, it suffices to consider $|s|\leq L/2$. Across this bipartition, the frozen state admits a Schmidt decomposition
\be \ket{\psi^s_L}=\sum_{i=1}^W\mu_i\ket{\psi^{s_i}_{L/2}}\otimes\ket{\psi^{s'_i}_{L/2}}, \ee
where $\sum_i|\mu_i|^2=1$. Each pair $(s_i,s'_i)$ corresponds to a way of partitioning the string $s$ compatible with the dynamics. Concretely, these strings are obtained by truncating $s$,
\be s_i:=s_{[1:x_i]}, s'_i:=s_{[x_i+1:|s|]}, \ee
for some allowed cut position $x_i$. The Schmidt decomposition thus corresponds to possible ways of distributing $s$ to the two subsystems. The precise Schmidt rank $W=|\{(s_i,s'_i)\}|$
depends on the constrained dynamics. For example, in the quantum fragmented $tJ_z$ model the Schmidt rank saturates the bound $W=|s|+1$, whereas in more strongly fragmented models such as the dipole-conserving model the Schmidt rank can be further reduced. The von Neumann entropy of subsystem $A$ is
\be S_A:=-\text{Tr}(\rho_A\ln\rho_A)=-\sum_i|\mu_i|^2\ln{|\mu_i|^2}. \ee
Using the fact that the entropy is maximized for a uniform distribution, we obtain the bound
\be S_A\leq \ln{W}\sim O(\ln{|s|}). \ee
When $|s|\sim O(1)$, the entropy satisfies an area-law scaling; whereas when $|s|\sim O(L)$, the entropy grows only logarithmically with system size $S_A\sim O(\ln(L))$.

We illustrate this bound with an explicit example in the TL model. Note that in this model, the truncated irreducible strings $(s_i,s'_i)$ always have the same parity with the corresponding subsystem sizes. Consider an entangled frozen state with irreducible string $s=(1010)$ on a chain of length $L=8$,
\be \ket{\psi^{(1010)}_8} :=\sqrt{\frac{1}{28}}\left(|10100000\ran-|10111000\ran+\dots\right). \ee
Its Schmidt decomposition across the bipartition at the middle of the chain takes the form
\bea \ket{\psi^{(1010)}_8} &=\sqrt{\frac{3}{14}}\ket{1010}\otimes\ket{\psi^{(\emptyset)}_4} + \sqrt{\frac{4}{7}}\ket{\psi^{(10)}_4}\otimes\ket{\psi^{(10)}_4} \\
&\quad + \sqrt{\frac{3}{14}}\ket{\psi^{(\emptyset)}_4}\otimes\ket{1010},
\eea
where
\bea \ket{\psi^{(\emptyset)}_4}&=\sqrt{\frac{1}{6}}(\ket{0000}-\ket{1100}-\ket{0110}-\ket{0011}\\
&\qquad +\ket{1111}-\ket{1001}), \\
\ket{\psi^{(10)}_4}&= \frac{1}{2}\left(\ket{1000}-\ket{1110}-\ket{1011}+\ket{0010}\right).
\eea
This state has bipartite entropy $S_A=0.98$, which is much smaller than the Page value $L\ln{2}/2\approx2.77$.

Finally, we address the question raised in the introduction. We have shown that the bipartite entanglement entropy of the entangled basis is at most logarithmic in system size. On the other hand, these basis states can nevertheless exhibit long-range correlations (e.g., GHZ-type frozen states discussed in Sec.~\ref{sec:verify}). As a result, the entangled basis of QF models cannot be mapped by any FDLU circuit either to a product-state basis underlying classical fragmentation, or to a typical eigenbasis of ergodic systems with volume-law entanglement. This entanglement structure provides a characteristic signature of quantum fragmentation in our construction.
% Finally, we comment on the entanglement structure of the entangled basis used in our construction. Since these states are composed of tensor products of dimers and frozen segments, their bipartite entanglement entropy is strongly constrained. If the bipartition cuts between two independent segments, the entropy vanishes. For long-range entangled frozen states labeled by the CF label $s$, the entanglement entropy is then bounded by $\ln{|s|}$. Intuitively, the Schmidt decomposition of the state corresponds to different ways of partitioning the string $s$ between the two subsystems, leading to a Schmidt rank bounded by $O(|s|)$. Consequently, the entangled frozen state obeys an area law when $|s|\sim O(1)$, while in the worst case $|s|\sim O(L)$ the entanglement entropy grows only logarithmically with system size. A detailed proof is given in Appendix~\ref{App: entanglement}. On the other hand, these basis states can nevertheless exhibit long-range correlations (e.g., GHZ-type frozen states discussed above). As a result, the entangled basis of QF models cannot be mapped by any finite-depth local unitary circuit either to a product-state basis underlying classical fragmentation or to a typical eigenbasis of ergodic systems with volume-law entanglement. This entanglement structure provides a characteristic signature of quantum fragmentation of our construction.

\section{2D Quantum Fragmentation}
In this section, we present an example of quantum fragmentation in 2D. Following the same strategy used in 1D, we study the quantum fragmented counterpart of the quad-flip model which generalizes PF dynamics to two dimensions \cite{Sthal2024Topo}. 

In the quad-flip model, we consider an $L\times L$ lattice where $N$ local degrees of freedom are located on the edges of the lattice. The dynamics preserves the 1-form symmetry charges defined on any path $\mcc$ on the lattice,
\be
\hat{Q}^{\a}_\mcc = \sum_{e_j\in \mcc} (-1)^j|\a\ran\lan\a|_{e_j}, \a=0,\dots,N-1,
\ee
where the edges follow the order $\{e_0,e_1,\dots\}$ on the path $\mcc$. We consider the source-free condition, under which $\hat{Q}^\alpha_{\partial \mcr} = 0$ for any region $\mcr$ on the lattice. This condition prohibits an odd number of spins of the same color on any face, and as a result, spins of the same color must form non-intersecting closed loops. The minimal local dynamics compatible with the symmetry involves simultaneously flipping four spins of the same color around a vertex $v$, i.e.
\be
H_{\text{quad}}=\sum_v\sum_{\a,\beta=0}^{N-1}\xi^{\a\beta}_v\hat{A}^{\a\beta}_v,
\ee
where $\xi_v^{\alpha\beta}$ are arbitrary constants and $\hat{A}^{\a\beta}_v=\prod_{e\in v}|\a\ran\lan\beta|_e$ flips the spins in state $\beta $ on the surrounding edges of the vertex $v$ to state $\alpha$. In the graphical representation for $N=3$, a state under the application of $\hat{A}^{\a\beta}_v$ at the center becomes
\be
\Bigg\lvert\kern0pt%
    \begin{tikzpicture}[baseline={(0,-0.12)}]%
        \draw [thick, RoyalBlue, rounded corners=0.65ex] (-3.6ex, 1.2ex)--++(2.4ex, 0)--++(0, 2.4ex);
        \draw [thick, RoyalBlue, rounded corners=0.65ex] (3.6ex, -1.2ex)--++(-2.4ex, 0)--++(0, -2.4ex);
        \draw [thick, ForestGreen, rounded corners=0.65ex] (-3.6ex, -1.2ex)--++(2.4ex, 0)--++(0, -2.4ex);
        \draw [thick, ForestGreen, rounded corners=0.65ex] (3.6ex, 1.2ex)--++(-2.4ex, 0)--++(0, 2.4ex);
        \draw [thick, BrickRed, rounded corners=0.65ex] (-1.2ex, 0)--++(0, 1.2ex)--++(2.4ex, 0)--++(0, -2.4ex)--++(-2.4ex, 0)--(-1.2ex, 0);
        \draw [semithick] (-2.4ex, -2.4ex) rectangle (2.4ex, 2.4ex);%
        \draw [semithick] (0, -2.4ex) rectangle (0, 2.4ex);%
        \draw [semithick] (-2.4ex, 0) rectangle (2.4ex, 0);%
        \fill [BrickRed] (-1.2ex, 0) circle(0.45ex);%
        \fill [BrickRed] (1.2ex, 0) circle(0.45ex);%
        \fill [BrickRed] (0, -1.2ex) circle(0.45ex);%
        \fill [BrickRed] (0, 1.2ex) circle(0.45ex);%
        \fill [RoyalBlue] (-1.2ex, 2.4ex) circle(0.45ex);%
        \fill [ForestGreen] (1.2ex, 2.4ex) circle(0.45ex);%
        \fill [RoyalBlue] (-2.4ex, 1.2ex) circle(0.45ex);%
        \fill [ForestGreen] (2.4ex, 1.2ex) circle(0.45ex);%
        \fill [ForestGreen] (-1.2ex, -2.4ex) circle(0.45ex);%
        \fill [RoyalBlue] (1.2ex, -2.4ex) circle(0.45ex);%
        \fill [ForestGreen] (-2.4ex, -1.2ex) circle(0.45ex);%
        \fill [RoyalBlue] (2.4ex, -1.2ex) circle(0.45ex);%
    \end{tikzpicture}\kern0pt\Bigg\rangle\to
\xi^{01}_v\Bigg\lvert\kern0pt%
    \begin{tikzpicture}[baseline={(0,-0.12)}]%
        \draw [thick, RoyalBlue, rounded corners=0.65ex] (-3.6ex, 1.2ex)--++(4.8ex, 0)--++(0, -4.8ex);
        \draw [thick, RoyalBlue, rounded corners=0.65ex] (3.6ex, -1.2ex)--++(-4.8ex, 0)--++(0, 4.8ex);
        \draw [thick, ForestGreen, rounded corners=0.65ex] (-3.6ex, -1.2ex)--++(2.4ex, 0)--++(0, -2.4ex);
        \draw [thick, ForestGreen, rounded corners=0.65ex] (3.6ex, 1.2ex)--++(-2.4ex, 0)--++(0, 2.4ex);
        \draw [semithick] (-2.4ex, -2.4ex) rectangle (2.4ex, 2.4ex);%
        \draw [semithick] (0, -2.4ex) rectangle (0, 2.4ex);%
        \draw [semithick] (-2.4ex, 0) rectangle (2.4ex, 0);%
        \fill [RoyalBlue] (-1.2ex, 0) circle(0.45ex);%
        \fill [RoyalBlue] (1.2ex, 0) circle(0.45ex);%
        \fill [RoyalBlue] (0, -1.2ex) circle(0.45ex);%
        \fill [RoyalBlue] (0, 1.2ex) circle(0.45ex);%
        \fill [RoyalBlue] (-1.2ex, 2.4ex) circle(0.45ex);%
        \fill [ForestGreen] (1.2ex, 2.4ex) circle(0.45ex);%
        \fill [RoyalBlue] (-2.4ex, 1.2ex) circle(0.45ex);%
        \fill [ForestGreen] (2.4ex, 1.2ex) circle(0.45ex);%
        \fill [ForestGreen] (-1.2ex, -2.4ex) circle(0.45ex);%
        \fill [RoyalBlue] (1.2ex, -2.4ex) circle(0.45ex);%
        \fill [ForestGreen] (-2.4ex, -1.2ex) circle(0.45ex);%
        \fill [RoyalBlue] (2.4ex, -1.2ex) circle(0.45ex);%
        \end{tikzpicture}\kern0pt\Bigg\rangle
+\xi^{11}_v\Bigg\lvert\kern0pt%
    \begin{tikzpicture}[baseline={(0,-0.12)}]%
        \draw [thick, RoyalBlue, rounded corners=0.65ex] (-3.6ex, 1.2ex)--++(2.4ex, 0)--++(0, 2.4ex);
        \draw [thick, RoyalBlue, rounded corners=0.65ex] (3.6ex, -1.2ex)--++(-2.4ex, 0)--++(0, -2.4ex);
        \draw [thick, ForestGreen, rounded corners=0.65ex] (-3.6ex, -1.2ex)--++(2.4ex, 0)--++(0, -2.4ex);
        \draw [thick, ForestGreen, rounded corners=0.65ex] (3.6ex, 1.2ex)--++(-2.4ex, 0)--++(0, 2.4ex);
        \draw [thick, BrickRed, rounded corners=0.65ex] (-1.2ex, 0)--++(0, 1.2ex)--++(2.4ex, 0)--++(0, -2.4ex)--++(-2.4ex, 0)--(-1.2ex, 0);
        \draw [semithick] (-2.4ex, -2.4ex) rectangle (2.4ex, 2.4ex);%
        \draw [semithick] (0, -2.4ex) rectangle (0, 2.4ex);%
        \draw [semithick] (-2.4ex, 0) rectangle (2.4ex, 0);%
        \fill [BrickRed] (-1.2ex, 0) circle(0.45ex);%
        \fill [BrickRed] (1.2ex, 0) circle(0.45ex);%
        \fill [BrickRed] (0, -1.2ex) circle(0.45ex);%
        \fill [BrickRed] (0, 1.2ex) circle(0.45ex);%
        \fill [RoyalBlue] (-1.2ex, 2.4ex) circle(0.45ex);%
        \fill [ForestGreen] (1.2ex, 2.4ex) circle(0.45ex);%
        \fill [RoyalBlue] (-2.4ex, 1.2ex) circle(0.45ex);%
        \fill [ForestGreen] (2.4ex, 1.2ex) circle(0.45ex);%
        \fill [ForestGreen] (-1.2ex, -2.4ex) circle(0.45ex);%
        \fill [RoyalBlue] (1.2ex, -2.4ex) circle(0.45ex);%
        \fill [ForestGreen] (-2.4ex, -1.2ex) circle(0.45ex);%
        \fill [RoyalBlue] (2.4ex, -1.2ex) circle(0.45ex);%
    \end{tikzpicture}\kern0pt\Bigg\rangle 
    +\xi^{12}_v\Bigg\lvert\kern0pt%
    \begin{tikzpicture}[baseline={(0,-0.12)}]%        
        \draw [thick, RoyalBlue, rounded corners=0.65ex] (-3.6ex, 1.2ex)--++(2.4ex, 0)--++(0, 2.4ex);
        \draw [thick, RoyalBlue, rounded corners=0.65ex] (3.6ex, -1.2ex)--++(-2.4ex, 0)--++(0, -2.4ex);
        \draw [thick, ForestGreen, rounded corners=0.65ex] (-3.6ex, -1.2ex)--++(4.8ex, 0)--++(0, 4.8ex);
        \draw [thick, ForestGreen, rounded corners=0.65ex] (3.6ex, 1.2ex)--++(-4.8ex, 0)--++(0, -4.8ex);
        \draw [semithick] (-2.4ex, -2.4ex) rectangle (2.4ex, 2.4ex);%
        \draw [semithick] (0, -2.4ex) rectangle (0, 2.4ex);%
        \draw [semithick] (-2.4ex, 0) rectangle (2.4ex, 0);%
        \fill [ForestGreen] (-1.2ex, 0) circle(0.45ex);%
        \fill [ForestGreen] (1.2ex, 0) circle(0.45ex);%
        \fill [ForestGreen] (0, -1.2ex) circle(0.45ex);%
        \fill [ForestGreen] (0, 1.2ex) circle(0.45ex);%
        \fill [RoyalBlue] (-1.2ex, 2.4ex) circle(0.45ex);%
        \fill [ForestGreen] (1.2ex, 2.4ex) circle(0.45ex);%
        \fill [RoyalBlue] (-2.4ex, 1.2ex) circle(0.45ex);%
        \fill [ForestGreen] (2.4ex, 1.2ex) circle(0.45ex);%
        \fill [ForestGreen] (-1.2ex, -2.4ex) circle(0.45ex);%
        \fill [RoyalBlue] (1.2ex, -2.4ex) circle(0.45ex);%
        \fill [ForestGreen] (-2.4ex, -1.2ex) circle(0.45ex);%
        \fill [RoyalBlue] (2.4ex, -1.2ex) circle(0.45ex);%
        \end{tikzpicture}\kern0pt\Bigg\rangle 
\ee
hence the name \textit{quad-flip} dynamics. Therefore, contractible loops can change color under the dynamics, whereas non-contractible loops that wrap around the system only undergo local fluctuations without altering their sequence. All states sharing the same configuration of non-contractible loops are dynamically connected and thus belong to the same Krylov sector. To identify the label of the sector, one can select a horizontal or vertical path across the system. The spin configuration along this path forms a 1D PF chain, from which one can extract the irreducible string to determine the horizontal or vertical label of the Krylov sector the state belongs to. More specifically, the 1D cut of a contractible loop appears as a pair of identical spins that can change color under the dynamics, while a horizontal (vertical) non-contractible loop intersects the vertical (horizontal) path only once, leaving an unpaired spin in the 1D cut. The only frozen states of the quad-flip model are thus the close-packed configurations of non-contractible loops arranged such that no two adjacent loops share the same color. 

When the quad-flip model becomes isotropic— i.e., when the matrix $\xi^{\alpha\beta}_v$ is the unit matrix (all entries one) — it exhibits quantum fragmentation, analogous to the TL model in 1D. For example, the state
\be
\Bigg\lvert\kern0pt%
    \begin{tikzpicture}[baseline={(0,-0.12)}]%
        \draw [thick, BrickRed, rounded corners=0.65ex] (-3.6ex, 2.4ex)--++(0,-1.2ex)--++(-1.2ex,0)--++(0,-2.4ex)--++(1.2ex,0)--++(0,-1.2ex);
        \draw [thick, ForestGreen, rounded corners=0.65ex] (-1.2ex, 2.4ex)--++(0,-1.2ex)--++(-2.4ex,0)--++(0,-2.4ex)--++(2.4ex,0)--++(0,-1.2ex);

        \draw [thick, ForestGreen, rounded corners=0.65ex] (1.2ex, 2.4ex)--++(0,-1.2ex)--++(2.4ex,0)--++(0,1.2ex);
        \draw [thick, ForestGreen, rounded corners=0.65ex] (1.2ex, -2.4ex)--++(0,1.2ex)--++(2.4ex,0)--++(0,-1.2ex);

        \draw [thick, BrickRed, dash pattern=on 3pt off 1.5pt,, rounded corners=0.65ex] (-1.4ex, 0.4ex)--++(0, 1ex)--++(1ex, 0);
        \draw [thick, BrickRed, dash pattern=on 3pt off 1.5pt,, rounded corners=0.65ex] (0.4ex, 1.4ex)--++(1ex, 0)--++(0, -1ex);
        \draw [thick, BrickRed, dash pattern=on 3pt off 1.5pt,, rounded corners=0.65ex] (1.4ex, -0.4ex)--++(0, -1ex)--++(-1ex,0);
        \draw [thick, BrickRed, dash pattern=on 3pt off 1.5pt,, rounded corners=0.65ex] (-0.4ex, -1.4ex)--++(-1ex,0)--++(0,1ex);
        \draw [thick, RoyalBlue, dash pattern=on 3pt off 1pt, rounded corners=0.65ex] (-1.1ex, 0.2ex)--++(0, 0.9ex)--++(0.9ex, 0);
        \draw [thick, RoyalBlue, dash pattern=on 3pt off 1pt, rounded corners=0.65ex] (0.2ex, 1.1ex)--++(0.9ex, 0)--++(0, -0.9ex);
        \draw [thick, RoyalBlue, dash pattern=on 3pt off 1pt, rounded corners=0.65ex] (1.1ex, -0.2ex)--++(0, -0.9ex)--++(-0.9ex,0);
        \draw [thick, RoyalBlue, dash pattern=on 3pt off 1pt, rounded corners=0.65ex] (-0.2ex, -1.1ex)--++(-0.9ex,0)--++(0,0.9ex);
        \draw [semithick] (-2.4ex, -2.4ex) rectangle (2.4ex, 2.4ex);%
        \draw [semithick] (0, -2.4ex) rectangle (0, 2.4ex);%
        \draw [semithick] (-2.4ex, 0) rectangle (2.4ex, 0);%
        \draw [semithick] (-4.8ex,2.4ex)--++(2.4ex,0);
        \draw [semithick] (-4.8ex,-2.4ex)--++(2.4ex,0);
        \draw [semithick] (2.4ex,2.4ex)--++(2.4ex,0);
        \draw [semithick] (2.4ex,-2.4ex)--++(2.4ex,0);
        \fill [black] (-1.2ex, 0) circle(0.45ex);%
        \fill [black] (1.2ex, 0) circle(0.45ex);%
        \fill [black] (0, -1.2ex) circle(0.45ex);%
        \fill [black] (0, 1.2ex) circle(0.45ex);%
        \fill [ForestGreen] (-1.2ex, 2.4ex) circle(0.45ex);%
        \fill [ForestGreen] (1.2ex, 2.4ex) circle(0.45ex);%
        \fill [ForestGreen] (-2.4ex, 1.2ex) circle(0.45ex);%
        \fill [ForestGreen] (2.4ex, 1.2ex) circle(0.45ex);%
        \fill [ForestGreen] (-1.2ex, -2.4ex) circle(0.45ex);%
        \fill [ForestGreen] (1.2ex, -2.4ex) circle(0.45ex);%
        \fill [ForestGreen] (-2.4ex, -1.2ex) circle(0.45ex);%
        \fill [ForestGreen] (2.4ex, -1.2ex) circle(0.45ex);%
        \fill [BrickRed] (-3.6ex, 2.4ex) circle(0.45ex);%
        \fill [BrickRed] (-3.6ex, -2.4ex) circle(0.45ex);%
        \fill [ForestGreen] (3.6ex, 2.4ex) circle(0.45ex);%
        \fill [ForestGreen] (3.6ex, -2.4ex) circle(0.45ex);%
    \end{tikzpicture}\kern0pt\Bigg\rangle \sim
    \Bigg\lvert\kern0pt%
    \begin{tikzpicture}[baseline={(0,-0.12)}]%
        \draw [thick, BrickRed, rounded corners=0.65ex] (-3.6ex, 2.4ex)--++(0,-1.2ex)--++(-1.2ex,0)--++(0,-2.4ex)--++(1.2ex,0)--++(0,-1.2ex);
        \draw [thick, ForestGreen, rounded corners=0.65ex] (-1.2ex, 2.4ex)--++(0,-1.2ex)--++(-2.4ex,0)--++(0,-2.4ex)--++(2.4ex,0)--++(0,-1.2ex);

        \draw [thick, ForestGreen, rounded corners=0.65ex] (1.2ex, 2.4ex)--++(0,-1.2ex)--++(2.4ex,0)--++(0,1.2ex);
        \draw [thick, ForestGreen, rounded corners=0.65ex] (1.2ex, -2.4ex)--++(0,1.2ex)--++(2.4ex,0)--++(0,-1.2ex);

        \draw [thick, BrickRed, rounded corners=0.65ex] (-1.2ex, 0)--++(0, 1.2ex)--++(2.4ex, 0)--++(0, -2.4ex)--++(-2.4ex, 0)--(-1.2ex, 0);
        \draw [semithick] (-2.4ex, -2.4ex) rectangle (2.4ex, 2.4ex);%
        \draw [semithick] (0, -2.4ex) rectangle (0, 2.4ex);%
        \draw [semithick] (-2.4ex, 0) rectangle (2.4ex, 0);%
        \draw [semithick] (-4.8ex,2.4ex)--++(2.4ex,0);
        \draw [semithick] (-4.8ex,-2.4ex)--++(2.4ex,0);
        \draw [semithick] (2.4ex,2.4ex)--++(2.4ex,0);
        \draw [semithick] (2.4ex,-2.4ex)--++(2.4ex,0);
        \fill [BrickRed] (-1.2ex, 0) circle(0.45ex);%
        \fill [BrickRed] (1.2ex, 0) circle(0.45ex);%
        \fill [BrickRed] (0, -1.2ex) circle(0.45ex);%
        \fill [BrickRed] (0, 1.2ex) circle(0.45ex);%
        \fill [ForestGreen] (-1.2ex, 2.4ex) circle(0.45ex);%
        \fill [ForestGreen] (1.2ex, 2.4ex) circle(0.45ex);%
        \fill [ForestGreen] (-2.4ex, 1.2ex) circle(0.45ex);%
        \fill [ForestGreen] (2.4ex, 1.2ex) circle(0.45ex);%
        \fill [ForestGreen] (-1.2ex, -2.4ex) circle(0.45ex);%
        \fill [ForestGreen] (1.2ex, -2.4ex) circle(0.45ex);%
        \fill [ForestGreen] (-2.4ex, -1.2ex) circle(0.45ex);%
        \fill [ForestGreen] (2.4ex, -1.2ex) circle(0.45ex);%
        \fill [BrickRed] (-3.6ex, 2.4ex) circle(0.45ex);%
        \fill [BrickRed] (-3.6ex, -2.4ex) circle(0.45ex);%
        \fill [ForestGreen] (3.6ex, 2.4ex) circle(0.45ex);%
        \fill [ForestGreen] (3.6ex, -2.4ex) circle(0.45ex);%
    \end{tikzpicture}\kern0pt\Bigg\rangle -
    \Bigg\lvert\kern0pt%
    \begin{tikzpicture}[baseline={(0,-0.12)}]%
        \draw [thick, BrickRed, rounded corners=0.65ex] (-3.6ex, 2.4ex)--++(0,-1.2ex)--++(-1.2ex,0)--++(0,-2.4ex)--++(1.2ex,0)--++(0,-1.2ex);
        \draw [thick, ForestGreen, rounded corners=0.65ex] (-1.2ex, 2.4ex)--++(0,-1.2ex)--++(-2.4ex,0)--++(0,-2.4ex)--++(2.4ex,0)--++(0,-1.2ex);
        \draw [thick, ForestGreen, rounded corners=0.65ex] (1.2ex, 2.4ex)--++(0,-1.2ex)--++(2.4ex,0)--++(0,1.2ex);
        \draw [thick, ForestGreen, rounded corners=0.65ex] (1.2ex, -2.4ex)--++(0,1.2ex)--++(2.4ex,0)--++(0,-1.2ex);

        \draw [thick, RoyalBlue, rounded corners=0.65ex] (-1.2ex, 0)--++(0, 1.2ex)--++(2.4ex, 0)--++(0, -2.4ex)--++(-2.4ex, 0)--(-1.2ex, 0);
        \draw [semithick] (-2.4ex, -2.4ex) rectangle (2.4ex, 2.4ex);%
        \draw [semithick] (0, -2.4ex) rectangle (0, 2.4ex);%
        \draw [semithick] (-2.4ex, 0) rectangle (2.4ex, 0);%
        \draw [semithick] (-4.8ex,2.4ex)--++(2.4ex,0);
        \draw [semithick] (-4.8ex,-2.4ex)--++(2.4ex,0);
        \draw [semithick] (2.4ex,2.4ex)--++(2.4ex,0);
        \draw [semithick] (2.4ex,-2.4ex)--++(2.4ex,0);
        \fill [RoyalBlue] (-1.2ex, 0) circle(0.45ex);%
        \fill [RoyalBlue] (1.2ex, 0) circle(0.45ex);%
        \fill [RoyalBlue] (0, -1.2ex) circle(0.45ex);%
        \fill [RoyalBlue] (0, 1.2ex) circle(0.45ex);%
        \fill [ForestGreen] (-1.2ex, 2.4ex) circle(0.45ex);%
        \fill [ForestGreen] (1.2ex, 2.4ex) circle(0.45ex);%
        \fill [ForestGreen] (-2.4ex, 1.2ex) circle(0.45ex);%
        \fill [ForestGreen] (2.4ex, 1.2ex) circle(0.45ex);%
        \fill [ForestGreen] (-1.2ex, -2.4ex) circle(0.45ex);%
        \fill [ForestGreen] (1.2ex, -2.4ex) circle(0.45ex);%
        \fill [ForestGreen] (-2.4ex, -1.2ex) circle(0.45ex);%
        \fill [ForestGreen] (2.4ex, -1.2ex) circle(0.45ex);%
        \fill [BrickRed] (-3.6ex, 2.4ex) circle(0.45ex);%
        \fill [BrickRed] (-3.6ex, -2.4ex) circle(0.45ex);%
        \fill [ForestGreen] (3.6ex, 2.4ex) circle(0.45ex);%
        \fill [ForestGreen] (3.6ex, -2.4ex) circle(0.45ex);%
    \end{tikzpicture}\kern0pt\Bigg\rangle
\ee
is frozen under the dynamics.\footnote{We impose ``smooth'' boundary conditions along the horizontal edges and ``rough'' boundary conditions along the vertical edges, so that loops can terminate only on the smooth boundaries. As a result, it suffices to consider only the horizontal symmetry operators \cite{Sthal2024Topo}.} If we take a horizontal cut through the bulk of the system, we obtain the 1D entangled frozen state $|12\ran\otimes(|00\ran-|11\ran)/\sqrt{2}$. We can also have larger frozen areas, e.g., the state
\be
\Bigg\lvert\kern0pt%
    \begin{tikzpicture}[baseline={(0,-0.12)}]%
        \draw [semithick] (-2.4ex, -2.4ex) rectangle (4.8ex, 2.4ex);%
        \draw [semithick] (0, -2.4ex) rectangle (0, 2.4ex);%
        \draw [semithick] (-2.4ex, 0) rectangle (4.8ex, 0);%
        \draw [semithick] (2.4ex, -2.4ex) rectangle (2.4ex, 2.4ex);%
        \draw [thick, RoyalBlue] (1.2ex, -3.6ex) rectangle (1.2ex, 3.6ex);%
         % Step 1: Clip the region you want to fill
        \begin{scope}
            \clip[rounded corners=0.65ex](-1.2ex, 0) -- ++(0, 1.2ex)-- ++(4.8ex, 0) -- ++(0, -2.4ex)-- ++(-4.8ex, 0) -- cycle;

        % Step 2: Draw alternating colored diagonal lines inside the clip
        \foreach \i [evaluate=\i as \col using {ifthenelse(mod(round(\i*5),2)==0,"BrickRed","RoyalBlue")}] in {-4,-3.8,...,4} {
        \draw[thick, \col] (\i,-2.5) -- ++(3,3);}
        \end{scope}

        % Step 3: Draw the original dashed outline on top
        \draw [thick, Black, dash pattern=on 3pt off 1.5pt,, rounded corners=0.65ex] (-1.2ex, 0)--++(0, 1.2ex)--++(4.8ex, 0)--++(0, -2.4ex)--++(-4.8ex, 0)--cycle;

        \draw [thick, ForestGreen, rounded corners=0.65ex] (-3.6ex, -1.2ex)--++(2.4ex, 0)--++(0, -2.4ex);
        \draw [thick, ForestGreen, rounded corners=0.65ex] (-3.6ex, 1.2ex)--++(2.4ex, 0)--++(0, 2.4ex);
        \draw [thick, ForestGreen, rounded corners=0.65ex] (6ex, 1.2ex)--++(-2.4ex, 0)--++(0, 2.4ex);
        \draw [thick, ForestGreen, rounded corners=0.65ex] (6ex, -1.2ex)--++(-2.4ex, 0)--++(0, -2.4ex);

        \fill [Black] (-1.2ex, 0) circle(0.45ex);%
        \fill [Black] (0, 1.2ex) circle(0.45ex);%
        \fill [Black] (2.4ex, 1.2ex) circle(0.45ex);%
        \fill [Black] (0, -1.2ex) circle(0.45ex);%
        \fill [Black] (2.4ex, -1.2ex) circle(0.45ex);%
        \fill [Black] (3.6ex, 0) circle(0.45ex);%

        \fill [ForestGreen] (-1.2ex, -2.4ex) circle(0.45ex);%
        \fill [ForestGreen] (-2.4ex, -1.2ex) circle(0.45ex);%
        \fill [ForestGreen] (-1.2ex, 2.4ex) circle(0.45ex);%
        \fill [ForestGreen] (-2.4ex, 1.2ex) circle(0.45ex);%
        \fill [ForestGreen] (3.6ex, 2.4ex) circle(0.45ex);%
        \fill [ForestGreen] (4.8ex, 1.2ex) circle(0.45ex);%
        \fill [ForestGreen] (3.6ex, -2.4ex) circle(0.45ex);%
        \fill [ForestGreen] (4.8ex, -1.2ex) circle(0.45ex);%

        \fill [RoyalBlue] (1.2ex, 2.4ex) circle(0.45ex);%
        \fill [RoyalBlue] (1.2ex, -2.4ex) circle(0.45ex);%
    \end{tikzpicture}\kern0pt\Bigg\rangle \sim
    \Bigg\lvert\kern0pt%
    \begin{tikzpicture}[baseline={(0,-0.12)}]%
        \draw [semithick] (-2.4ex, -2.4ex) rectangle (4.8ex, 2.4ex);%
        \draw [semithick] (0, -2.4ex) rectangle (0, 2.4ex);%
        \draw [semithick] (-2.4ex, 0) rectangle (4.8ex, 0);%
        \draw [semithick] (2.4ex, -2.4ex) rectangle (2.4ex, 2.4ex);%

        \draw [thick, RoyalBlue, rounded corners=0.65ex] (-1.2ex, 0)--++(0, 1.2ex)--++(2.4ex, 0)--++(0, -2.4ex)--++(-2.4ex, 0)--cycle;
        \draw [thick, RoyalBlue, rounded corners=0.65ex] (1.2ex,3.6ex)--++(0,-2.4ex)--++(2.4ex,0)--++(0,-2.4ex)--++(-2.4ex, 0)--++(0,-2.4ex);

        \draw [thick, ForestGreen, rounded corners=0.65ex] (-3.6ex, -1.2ex)--++(2.4ex, 0)--++(0, -2.4ex);
        \draw [thick, ForestGreen, rounded corners=0.65ex] (-3.6ex, 1.2ex)--++(2.4ex, 0)--++(0, 2.4ex);
        \draw [thick, ForestGreen, rounded corners=0.65ex] (6ex, 1.2ex)--++(-2.4ex, 0)--++(0, 2.4ex);
        \draw [thick, ForestGreen, rounded corners=0.65ex] (6ex, -1.2ex)--++(-2.4ex, 0)--++(0, -2.4ex);

        \fill [RoyalBlue] (-1.2ex, 0) circle(0.45ex);%
        \fill [RoyalBlue] (0, 1.2ex) circle(0.45ex);%
        \fill [RoyalBlue] (2.4ex, 1.2ex) circle(0.45ex);%
        \fill [RoyalBlue] (0, -1.2ex) circle(0.45ex);%
        \fill [RoyalBlue] (2.4ex, -1.2ex) circle(0.45ex);%
        \fill [RoyalBlue] (3.6ex, 0) circle(0.45ex);%
        \fill [RoyalBlue] (1.2ex, 0) circle(0.45ex);%

        \fill [ForestGreen] (-1.2ex, -2.4ex) circle(0.45ex);%
        \fill [ForestGreen] (-2.4ex, -1.2ex) circle(0.45ex);%
        \fill [ForestGreen] (-1.2ex, 2.4ex) circle(0.45ex);%
        \fill [ForestGreen] (-2.4ex, 1.2ex) circle(0.45ex);%
        \fill [ForestGreen] (3.6ex, 2.4ex) circle(0.45ex);%
        \fill [ForestGreen] (4.8ex, 1.2ex) circle(0.45ex);%
        \fill [ForestGreen] (3.6ex, -2.4ex) circle(0.45ex);%
        \fill [ForestGreen] (4.8ex, -1.2ex) circle(0.45ex);%

        \fill [RoyalBlue] (1.2ex, 2.4ex) circle(0.45ex);%
        \fill [RoyalBlue] (1.2ex, -2.4ex) circle(0.45ex);%
    \end{tikzpicture}\kern0pt\Bigg\rangle -
    \Bigg\lvert\kern0pt%
    \begin{tikzpicture}[baseline={(0,-0.12)}]%
        \draw [semithick] (-2.4ex, -2.4ex) rectangle (4.8ex, 2.4ex);%
        \draw [semithick] (0, -2.4ex) rectangle (0, 2.4ex);%
        \draw [semithick] (-2.4ex, 0) rectangle (4.8ex, 0);%
        \draw [semithick] (2.4ex, -2.4ex) rectangle (2.4ex, 2.4ex);%

        \draw [thick, BrickRed, rounded corners=0.65ex] (-1.2ex, 0)--++(0, 1.2ex)--++(2.4ex, 0)--++(0, -2.4ex)--++(-2.4ex, 0)--cycle;
        \draw [thick, RoyalBlue, rounded corners=0.65ex] (1.2ex,3.6ex)--++(0,-2.4ex)--++(2.4ex,0)--++(0,-2.4ex)--++(-2.4ex, 0)--++(0,-2.4ex);

        \draw [thick, ForestGreen, rounded corners=0.65ex] (-3.6ex, -1.2ex)--++(2.4ex, 0)--++(0, -2.4ex);
        \draw [thick, ForestGreen, rounded corners=0.65ex] (-3.6ex, 1.2ex)--++(2.4ex, 0)--++(0, 2.4ex);
        \draw [thick, ForestGreen, rounded corners=0.65ex] (6ex, 1.2ex)--++(-2.4ex, 0)--++(0, 2.4ex);
        \draw [thick, ForestGreen, rounded corners=0.65ex] (6ex, -1.2ex)--++(-2.4ex, 0)--++(0, -2.4ex);

        \fill [BrickRed] (-1.2ex, 0) circle(0.45ex);%
        \fill [BrickRed] (0, 1.2ex) circle(0.45ex);%
        \fill [RoyalBlue] (2.4ex, 1.2ex) circle(0.45ex);%
        \fill [BrickRed] (0, -1.2ex) circle(0.45ex);%
        \fill [RoyalBlue] (2.4ex, -1.2ex) circle(0.45ex);%
        \fill [RoyalBlue] (3.6ex, 0) circle(0.45ex);%
        \fill [BrickRed] (1.2ex, 0) circle(0.45ex);%

        \fill [ForestGreen] (-1.2ex, -2.4ex) circle(0.45ex);%
        \fill [ForestGreen] (-2.4ex, -1.2ex) circle(0.45ex);%
        \fill [ForestGreen] (-1.2ex, 2.4ex) circle(0.45ex);%
        \fill [ForestGreen] (-2.4ex, 1.2ex) circle(0.45ex);%
        \fill [ForestGreen] (3.6ex, 2.4ex) circle(0.45ex);%
        \fill [ForestGreen] (4.8ex, 1.2ex) circle(0.45ex);%
        \fill [ForestGreen] (3.6ex, -2.4ex) circle(0.45ex);%
        \fill [ForestGreen] (4.8ex, -1.2ex) circle(0.45ex);%

        \fill [RoyalBlue] (1.2ex, 2.4ex) circle(0.45ex);%
        \fill [RoyalBlue] (1.2ex, -2.4ex) circle(0.45ex);%
    \end{tikzpicture}\kern0pt\Bigg\rangle -
        \Bigg\lvert\kern0pt%
    \begin{tikzpicture}[baseline={(0,-0.12)}]%
        \draw [semithick] (-2.4ex, -2.4ex) rectangle (4.8ex, 2.4ex);%
        \draw [semithick] (0, -2.4ex) rectangle (0, 2.4ex);%
        \draw [semithick] (-2.4ex, 0) rectangle (4.8ex, 0);%
        \draw [semithick] (2.4ex, -2.4ex) rectangle (2.4ex, 2.4ex);%

        \draw [thick, BrickRed, rounded corners=0.65ex] (1.2ex, 0)--++(0, 1.2ex)--++(2.4ex, 0)--++(0, -2.4ex)--++(-2.4ex, 0)--cycle;
        \draw [thick, RoyalBlue, rounded corners=0.65ex] (1.2ex,3.6ex)--++(0,-2.4ex)--++(-2.4ex,0)--++(0,-2.4ex)--++(2.4ex, 0)--++(0,-2.4ex);

        \draw [thick, ForestGreen, rounded corners=0.65ex] (-3.6ex, -1.2ex)--++(2.4ex, 0)--++(0, -2.4ex);
        \draw [thick, ForestGreen, rounded corners=0.65ex] (-3.6ex, 1.2ex)--++(2.4ex, 0)--++(0, 2.4ex);
        \draw [thick, ForestGreen, rounded corners=0.65ex] (6ex, 1.2ex)--++(-2.4ex, 0)--++(0, 2.4ex);
        \draw [thick, ForestGreen, rounded corners=0.65ex] (6ex, -1.2ex)--++(-2.4ex, 0)--++(0, -2.4ex);

        \fill [RoyalBlue] (-1.2ex, 0) circle(0.45ex);%
        \fill [RoyalBlue] (0, 1.2ex) circle(0.45ex);%
        \fill [BrickRed] (2.4ex, 1.2ex) circle(0.45ex);%
        \fill [RoyalBlue] (0, -1.2ex) circle(0.45ex);%
        \fill [BrickRed] (2.4ex, -1.2ex) circle(0.45ex);%
        \fill [BrickRed] (3.6ex, 0) circle(0.45ex);%
        \fill [BrickRed] (1.2ex, 0) circle(0.45ex);%

        \fill [ForestGreen] (-1.2ex, -2.4ex) circle(0.45ex);%
        \fill [ForestGreen] (-2.4ex, -1.2ex) circle(0.45ex);%
        \fill [ForestGreen] (-1.2ex, 2.4ex) circle(0.45ex);%
        \fill [ForestGreen] (-2.4ex, 1.2ex) circle(0.45ex);%
        \fill [ForestGreen] (3.6ex, 2.4ex) circle(0.45ex);%
        \fill [ForestGreen] (4.8ex, 1.2ex) circle(0.45ex);%
        \fill [ForestGreen] (3.6ex, -2.4ex) circle(0.45ex);%
        \fill [ForestGreen] (4.8ex, -1.2ex) circle(0.45ex);%

        \fill [RoyalBlue] (1.2ex, 2.4ex) circle(0.45ex);%
        \fill [RoyalBlue] (1.2ex, -2.4ex) circle(0.45ex);%
    \end{tikzpicture}\kern0pt\Bigg\rangle,
\ee
is annihilated by $\hat{A}_v$ which acts on the frozen region highlighted by the shaded area enclosed by a dashed line and filled with slanted lines, with the alternating colors representing the quantum states of the spins involved. Taking a horizontal cut through the \emph{bulk} of the frozen region gives the 1D three-qubit entangled frozen state $(|000\ran-|110\ran-|011\ran)/\sqrt{3}$. 

It is natural to study the dynamics of the isotropic quad-flip model in an entangled basis following similar logic in the 1D TL model. The corresponding ``dimer'' state in 2D is the equal-weight superposition of loops of all colors, e.g., for $N=3$
\be
    \Big\lvert\kern0pt%
    \begin{tikzpicture}[baseline={(0,-0.12)}]%
        \draw [thick, Black, rounded corners=0.65ex] (-1.2ex, 0)--++(0, 1.2ex)--++(2.4ex, 0)--++(0, -2.4ex)--++(-2.4ex, 0)--cycle;
        \draw [semithick] (0, -2.4ex) rectangle (0, 2.4ex);%
        \draw [semithick] (-2.4ex, 0) rectangle (2.4ex, 0);%
        \fill [black] (-1.2ex, 0) circle(0.45ex);%
        \fill [black] (1.2ex, 0) circle(0.45ex);%
        \fill [black] (0, -1.2ex) circle(0.45ex);%
        \fill [black] (0, 1.2ex) circle(0.45ex);%
    \end{tikzpicture}\kern0pt\Big\rangle :=
    \Big(\Big\lvert\kern0pt%
    \begin{tikzpicture}[baseline={(0,-0.12)}]%
        \draw [thick, RoyalBlue, rounded corners=0.65ex] (-1.2ex, 0)--++(0, 1.2ex)--++(2.4ex, 0)--++(0, -2.4ex)--++(-2.4ex, 0)--cycle;
        \draw [semithick] (0, -2.4ex) rectangle (0, 2.4ex);%
        \draw [semithick] (-2.4ex, 0) rectangle (2.4ex, 0);%
        \fill [RoyalBlue] (-1.2ex, 0) circle(0.45ex);%
        \fill [RoyalBlue] (1.2ex, 0) circle(0.45ex);%
        \fill [RoyalBlue] (0, -1.2ex) circle(0.45ex);%
        \fill [RoyalBlue] (0, 1.2ex) circle(0.45ex);%
    \end{tikzpicture}\kern0pt\Big\rangle +
    \Big\lvert\kern0pt%
    \begin{tikzpicture}[baseline={(0,-0.12)}]%
        \draw [thick, BrickRed, rounded corners=0.65ex] (-1.2ex, 0)--++(0, 1.2ex)--++(2.4ex, 0)--++(0, -2.4ex)--++(-2.4ex, 0)--cycle;
        \draw [semithick] (0, -2.4ex) rectangle (0, 2.4ex);%
        \draw [semithick] (-2.4ex, 0) rectangle (2.4ex, 0);%
        \fill [BrickRed] (-1.2ex, 0) circle(0.45ex);%
        \fill [BrickRed] (1.2ex, 0) circle(0.45ex);%
        \fill [BrickRed] (0, -1.2ex) circle(0.45ex);%
        \fill [BrickRed] (0, 1.2ex) circle(0.45ex);%
    \end{tikzpicture}\kern0pt\Big\rangle +
    \Big\lvert\kern0pt%
        \begin{tikzpicture}[baseline={(0,-0.12)}]%
        \draw [thick, ForestGreen, rounded corners=0.65ex] (-1.2ex, 0)--++(0, 1.2ex)--++(2.4ex, 0)--++(0, -2.4ex)--++(-2.4ex, 0)--cycle;
        \draw [semithick] (0, -2.4ex) rectangle (0, 2.4ex);%
        \draw [semithick] (-2.4ex, 0) rectangle (2.4ex, 0);%
        \fill [ForestGreen] (-1.2ex, 0) circle(0.45ex);%
        \fill [ForestGreen] (1.2ex, 0) circle(0.45ex);%
        \fill [ForestGreen] (0, -1.2ex) circle(0.45ex);%
        \fill [ForestGreen] (0, 1.2ex) circle(0.45ex);%
    \end{tikzpicture}\kern0pt\Big\rangle\Big)\Big/\sqrt{3}.
\ee
In addition, we consider loop-like GHZ states as new elements in the entangled basis in 2D. They appear naturally in the dynamics of states consisting of neighboring entangled frozen loops of the same colors, e.g., under the projection operator $\hat{A}_o$ applied at the center, the state
\bea
\Bigg\lvert\kern0pt%
    \begin{tikzpicture}[baseline={(0,-0.12)}]%
        \draw [semithick] (0, -4.8ex) rectangle (0, 4.8ex);%
        \draw [semithick] (-2.4ex, -2.4ex) rectangle (2.4ex, 2.4ex);%
        \draw [semithick] (-4.8ex, 0) rectangle (4.8ex, 0);%

        \draw [thick, BrickRed, dash pattern=on 3pt off 1.5pt,, rounded corners=0.65ex] (-1.3ex, 2.8ex)--++(0, 0.9ex)--++(2.6ex, 0)--++(0,-2.6ex)--++(-2.6ex, 0)--++(0,1.2ex);
        \draw [thick, RoyalBlue, dash pattern=on 3pt off 1pt, rounded corners=0.65ex] (-1ex, 2.5ex)--++(0, 0.9ex)--++(2ex, 0)--++(0, -2ex)--++(-2ex, 0)--++(0,0.9ex);

        \draw [thick, BrickRed, dash pattern=on 3pt off 1.5pt,, rounded corners=0.65ex] (-3.7ex, 0.4ex)--++(0, 0.9ex)--++(2.6ex, 0)--++(0,-2.6ex)--++(-2.6ex, 0)--++(0,1.2ex);
        \draw [thick, RoyalBlue, dash pattern=on 3pt off 1pt, rounded corners=0.65ex] (-3.4ex, 0.1ex)--++(0, 0.9ex)--++(2ex, 0)--++(0, -2ex)--++(-2ex, 0)--++(0,0.9ex);

        \draw [thick, BrickRed, dash pattern=on 3pt off 1.5pt,, rounded corners=0.65ex] (-1.3ex, -2ex)--++(0, 0.9ex)--++(2.6ex, 0)--++(0,-2.6ex)--++(-2.6ex, 0)--++(0,1.2ex);
        \draw [thick, RoyalBlue, dash pattern=on 3pt off 1pt, rounded corners=0.65ex] (-1ex, -2.3ex)--++(0, 0.9ex)--++(2ex, 0)--++(0, -2ex)--++(-2ex, 0)--++(0,0.9ex);

        \draw [thick, BrickRed, dash pattern=on 3pt off 1.5pt,, rounded corners=0.65ex] (1.1ex, 0.4ex)--++(0, 0.9ex)--++(2.6ex, 0)--++(0,-2.6ex)--++(-2.6ex, 0)--++(0,1.2ex);
        \draw [thick, RoyalBlue, dash pattern=on 3pt off 1pt, rounded corners=0.65ex] (1.4ex, 0.1ex)--++(0, 0.9ex)--++(2ex, 0)--++(0, -2ex)--++(-2ex, 0)--++(0,0.9ex);

        \fill [black] (-1.2ex, 0) circle(0.45ex);%
        \fill [black] (-2.4ex, 1.2ex) circle(0.45ex);%
        \fill [black] (-3.6ex, 0) circle(0.45ex);%
        \fill [black] (-2.4ex, -1.2ex) circle(0.45ex);%

        \fill [black] (1.2ex, 0) circle(0.45ex);%
        \fill [black] (2.4ex, 1.2ex) circle(0.45ex);%
        \fill [black] (3.6ex, 0) circle(0.45ex);%
        \fill [black] (2.4ex, -1.2ex) circle(0.45ex);%

        \fill [black] (1.2ex, 2.4ex) circle(0.45ex);%
        \fill [black] (0ex, 3.6ex) circle(0.45ex);%
        \fill [black] (-1.2ex, 2.4ex) circle(0.45ex);%
        \fill [black] (0, 1.2ex) circle(0.45ex);%

        \fill [black] (1.2ex, -2.4ex) circle(0.45ex);%
        \fill [black] (0ex, -3.6ex) circle(0.45ex);%
        \fill [black] (-1.2ex, -2.4ex) circle(0.45ex);%
        \fill [black] (0, -1.2ex) circle(0.45ex);%
    \end{tikzpicture}\kern0pt\Bigg\rangle
    &\xrightarrow{\hat{A}_o}\Bigg(\Bigg\lvert\kern0pt%
\begin{tikzpicture}[baseline={(0,-0.12)}]%
    \draw [semithick] (0, -4.8ex) rectangle (0, 4.8ex);%
    \draw [semithick] (-2.4ex, -2.4ex) rectangle (2.4ex, 2.4ex);%
    \draw [semithick] (-4.8ex, 0) rectangle (4.8ex, 0);%
    
    \draw [thick, RoyalBlue, rounded corners=0.65ex] (0ex, 3.6ex)--++(-1.2ex, 0)--++(0, -2.4ex)--++(-2.4ex, 0)--++(0, -2.4ex)--++(2.4ex, 0)--++(0, -2.4ex)--++(2.4ex, 0)--++(0, 2.4ex)--++(2.4ex, 0)--++(0, 2.4ex)--++(-2.4ex, 0)--++(0, 2.4ex)--++(-1.2ex, 0);

    \draw [thick, Black, rounded corners=0.65ex] (-1.2ex, 0)--++(0, 1.2ex)--++(2.4ex, 0)--++(0, -2.4ex)--++(-2.4ex, 0)--cycle;
        \fill [black] (-1.2ex, 0) circle(0.45ex);%
        \fill [RoyalBlue] (-2.4ex, 1.2ex) circle(0.45ex);%
        \fill [RoyalBlue] (-3.6ex, 0) circle(0.45ex);%
        \fill [RoyalBlue] (-2.4ex, -1.2ex) circle(0.45ex);%

        \fill [black] (1.2ex, 0) circle(0.45ex);%
        \fill [RoyalBlue] (2.4ex, 1.2ex) circle(0.45ex);%
        \fill [RoyalBlue] (3.6ex, 0) circle(0.45ex);%
        \fill [RoyalBlue] (2.4ex, -1.2ex) circle(0.45ex);%

        \fill [RoyalBlue] (1.2ex, 2.4ex) circle(0.45ex);%
        \fill [RoyalBlue] (0ex, 3.6ex) circle(0.45ex);%
        \fill [RoyalBlue] (-1.2ex, 2.4ex) circle(0.45ex);%
        \fill [black] (0, 1.2ex) circle(0.45ex);%

        \fill [RoyalBlue] (1.2ex, -2.4ex) circle(0.45ex);%
        \fill [RoyalBlue] (0ex, -3.6ex) circle(0.45ex);%
        \fill [RoyalBlue] (-1.2ex, -2.4ex) circle(0.45ex);%
        \fill [black] (0, -1.2ex) circle(0.45ex);%
\end{tikzpicture}\kern0pt\Bigg\rangle +
\Bigg\lvert\kern0pt%
\begin{tikzpicture}[baseline={(0,-0.12)}]%
    \draw [semithick] (0, -4.8ex) rectangle (0, 4.8ex);%
    \draw [semithick] (-2.4ex, -2.4ex) rectangle (2.4ex, 2.4ex);%
    \draw [semithick] (-4.8ex, 0) rectangle (4.8ex, 0);%
    
    \draw [thick, BrickRed, rounded corners=0.65ex] (0ex, 3.6ex)--++(-1.2ex, 0)--++(0, -2.4ex)--++(-2.4ex, 0)--++(0, -2.4ex)--++(2.4ex, 0)--++(0, -2.4ex)--++(2.4ex, 0)--++(0, 2.4ex)--++(2.4ex, 0)--++(0, 2.4ex)--++(-2.4ex, 0)--++(0, 2.4ex)--++(-1.2ex, 0);

    \draw [thick, Black, rounded corners=0.65ex] (-1.2ex, 0)--++(0, 1.2ex)--++(2.4ex, 0)--++(0, -2.4ex)--++(-2.4ex, 0)--cycle;
        \fill [black] (-1.2ex, 0) circle(0.45ex);%
        \fill [BrickRed] (-2.4ex, 1.2ex) circle(0.45ex);%
        \fill [BrickRed] (-3.6ex, 0) circle(0.45ex);%
        \fill [BrickRed] (-2.4ex, -1.2ex) circle(0.45ex);%

        \fill [black] (1.2ex, 0) circle(0.45ex);%
        \fill [BrickRed] (2.4ex, 1.2ex) circle(0.45ex);%
        \fill [BrickRed] (3.6ex, 0) circle(0.45ex);%
        \fill [BrickRed] (2.4ex, -1.2ex) circle(0.45ex);%

        \fill [BrickRed] (1.2ex, 2.4ex) circle(0.45ex);%
        \fill [BrickRed] (0ex, 3.6ex) circle(0.45ex);%
        \fill [BrickRed] (-1.2ex, 2.4ex) circle(0.45ex);%
        \fill [black] (0, 1.2ex) circle(0.45ex);%

        \fill [BrickRed] (1.2ex, -2.4ex) circle(0.45ex);%
        \fill [BrickRed] (0ex, -3.6ex) circle(0.45ex);%
        \fill [BrickRed] (-1.2ex, -2.4ex) circle(0.45ex);%
        \fill [black] (0, -1.2ex) circle(0.45ex);%
\end{tikzpicture}\kern0pt\Bigg\rangle\Bigg)/\sqrt{2} \\
&:=\Bigg\lvert\kern0pt%
\begin{tikzpicture}[baseline={(0,-0.12)}]%
    \draw [semithick] (0, -4.8ex) rectangle (0, 4.8ex);%
    \draw [semithick] (-2.4ex, -2.4ex) rectangle (2.4ex, 2.4ex);%
    \draw [semithick] (-4.8ex, 0) rectangle (4.8ex, 0);%
    \draw [thick, BrickRed, rounded corners=0.65ex] (0ex, 3.6ex)--++(-1.2ex, 0)--++(0, -1.2ex)
    (-2.4ex, 1.2ex)--++(-1.2ex, 0)--++(0, -1.2ex)
    (-2.4ex, -1.2ex)--++(1.2ex, 0)--++(0, -1.2ex)
    (0ex, -3.6ex)--++(1.2ex, 0)--++(0, 1.2ex)
    (2.4ex, -1.2ex)--++(1.2ex, 0)--++(0, 1.2ex)
    (2.4ex, 1.2ex)--++(-1.2ex, 0)--++(0, 1.2ex);
    \draw [thick, RoyalBlue, rounded corners=0.65ex] 
    (-1.2ex, 2.4ex)--++(0,-1.2ex)--++(-1.2ex,0)
    (-3.6ex, 0ex)--++(0,-1.2ex)--++(1.2ex,0)
    (-1.2ex, -2.4ex)--++(0,-1.2ex)--++(1.2ex,0)
    (1.2ex, -2.4ex)--++(0,1.2ex)--++(1.2ex,0)
    (3.6ex, 0ex)--++(0,1.2ex)--++(-1.2ex,0)
    (1.2ex, 2.4ex)--++(0,1.2ex)--++(-1.2ex,0);
        \draw [thick, Black, rounded corners=0.65ex] (-1.2ex, 0)--++(0, 1.2ex)--++(2.4ex, 0)--++(0, -2.4ex)--++(-2.4ex, 0)--cycle;
        \fill [black] (-1.2ex, 0) circle(0.45ex);%
        \fill [black] (-2.4ex, 1.2ex) circle(0.45ex);%
        \fill [black] (-3.6ex, 0) circle(0.45ex);%
        \fill [black] (-2.4ex, -1.2ex) circle(0.45ex);%

        \fill [black] (1.2ex, 0) circle(0.45ex);%
        \fill [black] (2.4ex, 1.2ex) circle(0.45ex);%
        \fill [black] (3.6ex, 0) circle(0.45ex);%
        \fill [black] (2.4ex, -1.2ex) circle(0.45ex);%

        \fill [black] (1.2ex, 2.4ex) circle(0.45ex);%
        \fill [black] (0ex, 3.6ex) circle(0.45ex);%
        \fill [black] (-1.2ex, 2.4ex) circle(0.45ex);%
        \fill [black] (0, 1.2ex) circle(0.45ex);%

        \fill [black] (1.2ex, -2.4ex) circle(0.45ex);%
        \fill [black] (0ex, -3.6ex) circle(0.45ex);%
        \fill [black] (-1.2ex, -2.4ex) circle(0.45ex);%
        \fill [black] (0, -1.2ex) circle(0.45ex);%    
\end{tikzpicture}\kern0pt\Bigg\rangle,
\label{Eq:GHZ loop}
\eea
where the solid line with alternating red and blue colors denotes the GHZ state which is the equal weight superposition of red (`1') and blue (`0') loops. Taking a 1D cut through the center, we obtain the corresponding (unnormalized) 1D dynamics 
\be (|00\ran-|11\ran)^{\otimes 2}\xrightarrow{\hat{P}_{2,3}} |0\:\tikz[baseline={([yshift=-0.7ex]current bounding box.center)}]{
\draw[thick,Black] (0,0)--(0.5,0);
\fill[black] (0,0) circle (2pt);
\fill[black] (0.5,0) circle (2pt);} \:0\ran+|1\:\tikz[baseline={([yshift=-0.7ex]current bounding box.center)}]{
\draw[thick,Black] (0,0)--(0.5,0);
\fill[black] (0,0) circle (2pt);
\fill[black] (0.5,0) circle (2pt);}\: 1\ran\ee
For non-contractible GHZ loops, however, the situation is more subtle. Their stability under the quad-flip dynamics depends on background configuration. For example, when the GHZ loop is surrounded by neighboring dimer loops in equal-weight superpositions of all colors—the quad-flip terms act identically on each constituent basis state of the GHZ loop, and the coherence of the superposition is preserved. For example, under the projector $\hat{A}_o$ at the center,
\be
\Bigg\lvert\kern0pt%
\begin{tikzpicture}[baseline={(0,-0.12)}]%
    \draw [semithick] (0, -4.8ex) rectangle (0, 2.4ex);%
    \draw [semithick] (-2.4ex, -2.4ex) rectangle (2.4ex, 2.4ex);%
    \draw [semithick] (-2.4ex, 0) rectangle (4.8ex, 0);%
    \draw [thick, Black, rounded corners=0.65ex] (-1.2ex, -2.4ex)--++(0, 1.2ex)--++(2.4ex, 0)--++(0, -2.4ex)--++(-2.4ex, 0)--cycle;
    \draw [thick, Black, rounded corners=0.65ex] (1.2ex, 0)--++(0, 1.2ex)--++(2.4ex, 0)--++(0, -2.4ex)--++(-2.4ex, 0)--cycle;
    \draw [thick, Black, rounded corners=0.65ex] (-3.6ex,1.2ex)--++(2.4ex,0)--++(0,2.4ex);
    \draw [thick, RoyalBlue, rounded corners=0.65ex] 
    (-3.6ex,-1.2ex)--++(1.2ex,0)
    (-1.2ex,0)--++(0,1.2ex)--++(1.2ex,0)
    (1.2ex,2.4ex)--++(0,1.2ex);
    \draw [thick, BrickRed, rounded corners=0.65ex]
    (-2.4ex,-1.2ex)--++(1.2ex,0)--++(0,1.2ex)
    (0,1.2ex)--++(1.2ex,0)--++(0,1.2ex);
    
        \fill [black] (-1.2ex, 0) circle(0.45ex);%
        \fill [black] (-2.4ex, 1.2ex) circle(0.45ex);%
        \fill [black] (-2.4ex, -1.2ex) circle(0.45ex);%

        \fill [black] (1.2ex, 0) circle(0.45ex);%
        \fill [black] (2.4ex, 1.2ex) circle(0.45ex);%
        \fill [black] (3.6ex, 0) circle(0.45ex);%
        \fill [black] (2.4ex, -1.2ex) circle(0.45ex);%

        \fill [black] (1.2ex, 2.4ex) circle(0.45ex);%
        \fill [black] (-1.2ex, 2.4ex) circle(0.45ex);%
        \fill [black] (0, 1.2ex) circle(0.45ex);%

        \fill [black] (1.2ex, -2.4ex) circle(0.45ex);%
        \fill [black] (0ex, -3.6ex) circle(0.45ex);%
        \fill [black] (-1.2ex, -2.4ex) circle(0.45ex);%
        \fill [black] (0, -1.2ex) circle(0.45ex);%    
\end{tikzpicture}\kern0pt\Bigg\rangle
\xrightarrow{\hat{A}_o} \Bigg\lvert\kern0pt%
\begin{tikzpicture}[baseline={(0,-0.12)}]%
    \draw [semithick] (0, -4.8ex) rectangle (0, 2.4ex);%
    \draw [semithick] (-2.4ex, -2.4ex) rectangle (2.4ex, 2.4ex);%
    \draw [semithick] (-2.4ex, 0) rectangle (4.8ex, 0);%
    \draw [thick, Black, rounded corners=0.65ex] (-3.6ex,1.2ex)--++(2.4ex,0)--++(0,2.4ex);
    \draw [thick, Black, rounded corners=0.65ex] (-1.2ex, 0)--++(0, 1.2ex)--++(2.4ex, 0)--++(0, -2.4ex)--++(-2.4ex, 0)--cycle;
    \draw [thick, RoyalBlue, rounded corners=0.65ex] 
    (-3.6ex,-1.2ex)--++(1.2ex,0)
    (-1.2ex,-2.4ex)--++(0,-1.2ex)--++(1.2ex,0)
    (1.2ex,-2.4ex)--++(0,1.2ex)--++(1.2ex,0)
    (3.6ex,0)--++(0,1.2ex)--++(-1.2ex,0)
    (1.2ex,2.4ex)--++(0,1.2ex);
    \draw [thick, BrickRed, rounded corners=0.65ex]
    (-2.4ex,-1.2ex)--++(1.2ex,0)--++(0,-1.2ex)
    (0,-3.6ex)--++(1.2ex,0)--++(0,1.2ex)
    (2.4ex,-1.2ex)--++(1.2ex,0)--++(0,1.2ex)
    (2.4ex,1.2ex)--++(-1.2ex,0)--++(0,1.2ex);

        \fill [black] (-1.2ex, 0) circle(0.45ex);%
        \fill [black] (-2.4ex, 1.2ex) circle(0.45ex);%
        \fill [black] (-2.4ex, -1.2ex) circle(0.45ex);%

        \fill [black] (1.2ex, 0) circle(0.45ex);%
        \fill [black] (2.4ex, 1.2ex) circle(0.45ex);%
        \fill [black] (3.6ex, 0) circle(0.45ex);%
        \fill [black] (2.4ex, -1.2ex) circle(0.45ex);%

        \fill [black] (1.2ex, 2.4ex) circle(0.45ex);%
        \fill [black] (-1.2ex, 2.4ex) circle(0.45ex);%
        \fill [black] (0, 1.2ex) circle(0.45ex);%

        \fill [black] (1.2ex, -2.4ex) circle(0.45ex);%
        \fill [black] (0ex, -3.6ex) circle(0.45ex);%
        \fill [black] (-1.2ex, -2.4ex) circle(0.45ex);%
        \fill [black] (0, -1.2ex) circle(0.45ex);%    
    \end{tikzpicture}\kern0pt\Bigg\rangle.
\ee
In this case, the loop only deforms locally under the action of projection operators. By contrast, in a single-colored background, the quad-flip terms act asymmetrically on the different constituent states, which leads to dephasing and eventual loss of coherence. Thus, unlike in the classically fragmented quad-flip model where non-contractible single-colored loops are strictly conserved, in the isotropic quantum fragmented model non-contractible GHZ loops are not absolutely protected but can nevertheless serve as long-lived, symmetry-protected carriers of entanglement. 

The above discussion makes clear that QF can arise in dimensions greater than one, and that new phenomena can be supported in a higher-dimensional context, e.g., QF based on entangled superpositions of loops, contractible or otherwise. An exhaustive characterization of 2D QF, analogous to the treatment of 1D QF in preceding sections, is beyond the scope of the present work. 

\section{Discussion}
In this work, we have developed a systematic framework for constructing quantum fragmented (QF) models. Our approach is based on a `Rokhsar-Kivelson' (RK) type protocol that promotes classically fragmented (CF) models to frustration free QF models built out of local projectors. The same construction can even be applied to non-fragmented models, which we have illustrated by promoting the transverse-field Ising model to a QF model. In one dimension, we have introduced an (overcomplete) labeling scheme for the resulting Krylov sectors by constructing an entangled basis, and demonstrated how it can be used to count sectors. We also provide an experimental verification of QF, assuming the ability to prepare specific initial states and perform tomography on few-site reduced density matrices. Finally, the protocol is not restricted to 1D: as proof of principle we have constructed a 2D QF model by lifting the quad-flip Hamiltonian. A complete characterization of QF beyond one dimension is left for future work.

A central outcome of our construction is the nontrivial entanglement structure of the resulting entangled basis underlying QF models. As we have shown, it consists of tensor products of dimers and frozen segments, which strongly limits the growth of bipartite entanglement. In particular, while generic cuts yield only $O(1)$ entropy, the most entangled states arise from non-separable frozen states and exhibit at most logarithmic scaling with system size. Remarkably, this sub-volume-law entanglement coexists with long-range quantum correlations. This combination sharply distinguishes QF from both classical fragmentation, where the basis is unentangled, and from generic ergodic systems trivially ``fragmented'' in its eigenbasis, which typically obey volume-law entanglement. Consequently, we formalize the notion of QF by requiring that it cannot be reduced to either limit via FDLU circuits.

It is natural to compare QF to anomalous symmetries (especially in 1D), in the sense of symmetries that do not admit short-range entangled (SRE) symmetric states. Unlike ordinary onsite symmetries (or non-onsite but “onsiteable” ones), the symmetry sectors of such anomalous symmetries cannot be spanned by SRE basis states~\cite{chen2011two, levin2012braiding, zhang2024long}. 
In this sense, anomalous symmetry is a close structural cousin of QF: both obstruct any reduction to a purely classical, product-state description by finite-depth local unitaries. However, the two notions are not equivalent, since anomalous symmetry does not by itself imply exponentially many dynamically disconnected Krylov sectors or non-ergodic dynamics. Clarifying the relation between anomalous symmetries and quantum fragmentation is an interesting direction for future work.

%Beyond existence, we showed how QF can be \emph{experimentally diagnosed}. Starting from a minimal non‑separable entangled frozen state, we proved that every $(k+1)$‑site reduced density matrix (RDM) contains a fixed entangled component aligned with the local projector kernel. In the TL model (e.g., $N=2$), this takes the form of a robust Bell‑pair contribution to any nearest‑neighbor RDM, providing a concrete, two‑site tomographic signature that distinguishes QF from CF while remaining accessible to current platforms.

%We then study the fragmented dynamics by constructing a labeling scheme for Krylov sectors in QF models. For the TL model we introduced an entangled basis made from non‑crossing dimers interleaved with frozen segments. This yields a description in which sector dynamics reduces to rearrangements/fusion of dimers and frozen segments. We further count the Krylov sectors and found a tight upperbound that, within each CF Krylov sector, the QF structure exhibits an exponential proliferation of sectors, reflecting the additional entangled block-diagonal structure invisible to any classical description. Although not explicitly illustrated, the labeling scheme for generic QF models should follow similar principles.

%Finally, we pushed QF beyond one dimension by constructing a \emph{two‑dimensional} example: the isotropic quad‑flip model with a source‑free 1‑form symmetry...

An open issue is how robust QF remains under perturbations that relax the idealized constraints of the model. Related questions have been explored in Refs.~\cite{han2026HSF,wang2025slowtherm,Li2025dynamics}. It was shown that 1D QF models of the type constructed here remain non-ergodic and fail to thermalize even when locally coupled to sparse single-site impurities (see Appendix F of Ref.~\cite{han2026HSF}). This robustness originates from exponentially large spectral degeneracies, which are stable against local perturbations and lead to persistent memory of initial conditions at arbitrarily long times. These results indicate that quantum fragmentation is not a fine-tuned Hamiltonian construction, but a robust dynamical phenomenon. Exploring potential applications, such as robust quantum memories, remains an interesting direction for future work~\cite{iadecola2025symmetryfragmentation, IadecolaNandkishore}.

Finally, the framework presented here is only one systematic route to constructing QF models. Alternative mechanisms that fall outside of this scheme are known to exist~\cite{Brighi2023quantumEast,balducci2025deephilbertspacealltoall}, and mapping out such possibilities is a natural target for future investigation. Recent independent work~\cite{zhou2026quantumhilbertspacefragmentation} further illustrates the richness of this direction.

\begin{acknowledgments}
We thank Zihan Zhou, Tian-Hua Yang, Huanqian Loh, Charles Stahl, Pablo Sala, Carolyn Zhang, Ruochen Ma and Thomas Iadecola for insightful discussions. This work was supported by AFOSR grant FA9550-20-1-0222 (YH, OH, RN), FA9550-24-1-0120 (YH), National Science Foundation grant GCR-2428487 (AK), and in part by the National Science Foundation grant NSF PHY-2309135 to the Kavli Institute for Theoretical Physics (KITP).
\end{acknowledgments}

\appendix
\section{Ground space of the QF promoted generalized transverse-field Ising model}\label{App: XX}
In this appendix, we prove that the ground space $\mcg_L$ of $H^{\text{QF}}_{\text{TFIM}}$ given in Eq.~\eqref{eq: QF_XX} has dimension
\be\label{eq: XX_dim} |\mcg_L| = N(N-1)^{L-1}.\ee
The ground space is defined as the subspace of states annihilated by every projector
\be \mcg_L=\bigcap_{i,a} \ker \Pi^a_{i,i+1} \ee

We determine $\mcg_L$ inductively. Given $\mcg_{L-1}$, the constraint on bond $(1,2)$ requires
\be \mcg_L=\ker(\sum_{a=0}^{N-1}\Pi^a_{1,2}:\mch\otimes\mcg_{L-1}\to \sum_{a=0}^{N-1} \lambda_a|\psi^a\ran\otimes|v_a\ran)\ee
where $\mch$ is the onsite Hilbert space and $|v_a\ran\in\mcg_{L-2}$.
In order to apply the rank-nullity theorem, we compute the dimension of the image. 

Any basis state in the domain can be expressed as 
\be |\xi\ran=|n\ran\otimes|\phi\ran\in \mch\otimes\mcg_{L-1}, \ee
where 
\be |\phi\ran := \sum_{b=0}^{N-1} \mu_b|b\ran\otimes |v_b\ran, |v_b\ran\in\mcg_{L-2}, \ee
and the coefficients satisfy the constraint $\sum_a \Pi^a_{2,3}|\phi\ran=0$. Applying the projectors on bond $(1,2)$, 
\bea \sum_a\Pi^a_{1,2}|\xi\ran &= \sum_{a,b}\mu_b\lan\psi^a|nb\ran|\psi^a\ran|v_b\ran \\
&= \sum_b \mu_b|\psi^{b-n}\ran|v_b\ran, \eea
where the last equality follows from the fact that $\lan\psi^a|nb\ran=1$ if and only if $b-n (\text{mod } N)=a$, and is zero otherwise. The dependence on $n$ only enters through the shift $b\mapsto b-n$, which is a permutation of the orbit indices. Therefore varying $n$ does not generate new subspaces in the image, the ensemble of states
\be \left\{\sum_b \mu_b|\psi^{b-n}\ran|v_b\ran, n=0,1,\dots, N-1\right\}\ee
spans the same space as the ensemble with $n=0$. 

Hence the dimension of the image is simply the number of independent choices of $|\phi\ran\in\mcg_{L-1}$, i.e.,
\be \dim\text{Im} = |\mcg_{L-1}|. \ee
We thus have the recurrence relation
\be |\mcg_L|=\dim[\mch\otimes\mcg_{L-1}]-\dim\text{Im} = (N-1)|\mcg_{L-1}|. \ee
Given the initial conditions $|\mcg_0|=1$ and $|\mcg_1|=N$, the solution is exactly Eq.~\eqref{eq: XX_dim}.
\qed

\section{Entangled frozen states in other quantum fragmented models}\label{Appendix: other QF models}
In this Appendix, we illustrate the construction of entangled frozen states in models with QF, other than the Temperley-Lieb (TL) model.

Recall that in the TL model, entangled frozen states may be separable, and their structure is characterized not only by the CF label but also by the lengths of the constituent frozen segments. However, in the quantum fragmented $tJz$ model defined in Eq.~\eqref{eq: QF_tJz}, there are no separable entangled frozen states, since spins can move freely provided there are vacant sites. As a result, each entangled frozen state is uniquely determined by a CF label and takes the form of an equal-weight superposition of all configurations sharing the same spin pattern. For example, when $L=3$, the frozen state labeled by $\ua$ is
\be \ket{\psi^{(\ua)}_{\text{froz}}}=\frac{1}{\sqrt{3}}(\ket{\ua00}+\ket{0\ua 0}+\ket{00\ua}). \ee

In the range-three charge-dipole model defined in Eq.~\eqref{eq:H^QF_3}, the dynamics preserves not only the total charge and total dipole moment, but also the pattern of “defects,” defined as spins that are identical to their left nearest occupied neighbors~\cite{Rakovszky2020localization}. Configurations with the same defect pattern further fragment into smaller subsectors, since the dipole moment $P_k$ between the $k$-th and $(k+1)$-th defects is separately conserved. In contrast to the $tJ_z$ model, defects here are constrained by the dipole moment conservation. This allows entangled frozen segments to be separated by product frozen segments such as $|\cdots +++\cdots\ran$, but in this case, all structural information is already encoded in the corresponding CF label. Consequently, each entangled frozen state is given by an equal-weight superposition of all configurations sharing the same CF label, with relative phases determined by their distance on the connectivity graph. For example, for $L=4$, an entangled frozen state labeled by an empty defect string, total charge $Q=0$, and dipole moment $P_0=2$, is
\bea |\psi^{(\emptyset,Q=0,P_0=2)}_{\text{froz}}\ran=
\frac{1}{\sqrt{3}}(\spacedket{- + - +}-\spacedket{0 - 0 +}-\spacedket{- 0 + 0}).
\eea

\section{Linear-independence and completeness of the entangled basis of the Temperley-Lieb model with \texorpdfstring{$\boldsymbol{N=2$}}{N=2}}\label{app:TL N=2}
In this appendix, we show that for the Temperley-Lieb (TL) model with local Hilbert space dimension $N=2$, any state can be spanned by the basis states of the form
\be |\psi\ran=|\psi_{\text{dimer}}\ran\otimes|\psi_{\text{{froz}}}\ran, \ee
where $|\psi_{\text{dimer}}\ran = \prod_{l=1}^{N_d}|\tikz[]{\fill[black] (0,0) circle (2pt);
\fill[black] (0.5,0) circle (2pt);
\draw[thick,black] (0,0)--(0.5,0);}\ran_{j_l,j_{l}+1}$ is the tensor product of $N_d$ non-crossing dimers defined as
\begin{equation}
    |\tikz[]{\fill[black] (0,0) circle (2pt);
\fill[black] (0.5,0) circle (2pt);
\draw[thick,black] (0,0)--(0.5,0);}\ran_{i,i+1}:=\frac{1}{\sqrt{N}}\sum_{a=1}^N|aa\ran_{i,i+1}.
\end{equation}
$|\psi_{\text{froz}}\ran$ are the (entangled) frozen segments. When $N=2$, an orthonormal basis for $|\psi_{\text{froz}}\ran$ within a frozen region of length $d$ consists of $d+1$ frozen states $|\psi^{(s),d}\ran$ with $s=\emptyset, (0101\dots)$ or $(1010\dots)$. To visualize these states, we can assign $N$ different colors to $N$ degrees of freedom per site, e.g., when $N=2$, we represent $|0\ran=|\textcolor{NavyBlue}{\bullet}\ran$ and $|1\ran=|\textcolor{BrickRed}{\bullet}\ran$. We use dashed lines with the corresponding colors involved to represent the entangled frozen states. For example, for $L=3$,
\begin{equation}
|\psi^{(0)}_3\ran=|000\ran-|110\ran-|011\ran=
|\tikz[baseline={([yshift=-1.5ex]current bounding box.center)}]{
\draw[dash pattern=on 3pt off 1.5pt,thick,NavyBlue] (0.06,0.035)--(0.54,0.035);
\draw[dash pattern=on 3pt off 1.5pt,thick,BrickRed] (0.06,-0.035)--(0.54,-0.035);
\draw[dash pattern=on 3pt off 1.5pt,thick,NavyBlue] (0.66,0.035)--(1.14,0.035);
\draw[dash pattern=on 3pt off 1.5pt,thick,BrickRed] (0.66,-0.035)--(1.14,-0.035);
\fill[black] (0,0) circle (2pt);
\fill[black] (0.6,0) circle (2pt);
\fill[black] (1.2,0) circle (2pt);
\node[] at (1.5,0.15) {\scriptsize $(0)$};
}\ran.
\end{equation}

We can define the projector onto the singlet state on sites $i$ and $i+1$
\begin{equation}
    \hat{P}_{i,i+1}=|\tikz[]{\fill[black] (0,0) circle (2pt);
\fill[black] (0.5,0) circle (2pt);
\draw[thick,black] (0,0)--(0.5,0);}\ran\lan\tikz[]{\fill[black] (0,0) circle (2pt);
\fill[black] (0.5,0) circle (2pt);
\draw[thick,black] (0,0)--(0.5,0);}|_{i,i+1},
\end{equation}
such that $H_{\text{TL}}=\sum_{i=1}^{L-1}J_i\hat{P}_{i,i+1}$. Then, any two states that differ on at least one bond, where one bond is occupied by a dimer and the other is annihilated by $\hat{P}_{i,i+1}$, are orthogonal. When these states have dimers on neighboring bonds, they are not orthogonal anymore, for example, $\lan\tikz[]{\fill[black] (0,0) circle (2pt);
\fill[black] (0.5,0) circle (2pt);
\fill[black] (1,0) circle (2pt);
\draw[thick,black] (0,0)--(0.5,0);}
|\tikz[]{\fill[black] (0,0) circle (2pt);
\fill[black] (0.5,0) circle (2pt);
\fill[black] (1,0) circle (2pt);
\draw[thick,black] (0.5,0)--(1,0);}\ran\neq 0$. However, the basis is still linearly-independent. We use induction to prove the linear-independence of the basis for $N=2$. First, let us explicitly check the basis for small $L$. When $L=2$, the basis 
\begin{equation}
\begin{aligned}
    \Omega(2)&=\left\{|\tikz[]{\fill[black] (0,0) circle (2pt);
\fill[black] (0.5,0) circle (2pt);
\draw[thick,black] (0,0)--(0.5,0);}\ran,
|\tikz[]{\fill[NavyBlue] (0,0) circle (2pt);
\fill[BrickRed] (0.5,0) circle (2pt);}\ran,
|\tikz[]{\fill[BrickRed] (0,0) circle (2pt);
\fill[NavyBlue] (0.5,0) circle (2pt);}\ran,
|\tikz[]{\draw[dash pattern=on 3pt off 1.5pt,thick,NavyBlue] (0.06,0.035)--(0.54,0.035);
\draw[dash pattern=on 3pt off 1.5pt,thick,BrickRed] (0.06,-0.035)--(0.54,-0.035);
\fill[black] (0,0) circle (2pt);
\fill[black] (0.5,0) circle (2pt);}\ran\right\}\\
    &=\left\{\frac{|00\ran+|11\ran}{\sqrt{2}},|01\ran,|10\ran,\frac{|00\ran-|11\ran}{\sqrt{2}}\right\}
\end{aligned}
\end{equation}
is clearly linearly-independent because these states are mutually orthogonal. When $L=3$, 
\begin{widetext}
\begin{equation}
% \begin{aligned}
    \Omega(3)=\{|\tikz[]{\fill[black] (0,0) circle (2pt);
\fill[black] (0.5,0) circle (2pt);
\draw[thick,black] (0,0)--(0.5,0);}\ran\otimes|\tikz[]{
\fill[NavyBlue] (0,0) circle (2pt);}\ran/|\tikz[]{
\fill[BrickRed] (0,0) circle (2pt);}\ran, \enspace
|\tikz[]{
\fill[NavyBlue] (0,0) circle (2pt);}\ran/|\tikz[]{
\fill[BrickRed] (0,0) circle (2pt);}\ran\otimes
|\tikz[]{\fill[black] (0,0) circle (2pt);
\fill[black] (0.5,0) circle (2pt);
\draw[thick,black] (0.5,0)--(0,0);}\ran, \enspace
|\tikz[]{\fill[NavyBlue] (0,0) circle (2pt);
\fill[BrickRed] (0.5,0) circle (2pt);
\fill[NavyBlue] (1,0) circle (2pt);}\ran, \enspace
|\tikz[]{\fill[BrickRed] (0,0) circle (2pt);
\fill[NavyBlue] (0.5,0) circle (2pt);
\fill[BrickRed] (1,0) circle (2pt);}\ran, \enspace
|\tikz[baseline={([yshift=-.6ex]current bounding box.center)}]{
\draw[dash pattern=on 3pt off 1.5pt,thick,NavyBlue] (0.06,0.035)--(0.54,0.035);
\draw[dash pattern=on 3pt off 1.5pt,thick,BrickRed] (0.06,-0.035)--(0.54,-0.035);
\draw[dash pattern=on 3pt off 1.5pt,thick,NavyBlue] (0.56,0.035)--(1,0.035);
\draw[dash pattern=on 3pt off 1.5pt,thick,BrickRed] (0.56,-0.035)--(1,-0.035);
\fill[black] (0,0) circle (2pt);
\fill[black] (0.5,0) circle (2pt);
\fill[black] (1,0) circle (2pt); 
% \node[] at (1.3,-0.15) {\scriptsize +};
\node[] at (1.5,0.1) {\scriptsize $(0)/(1)$};
}\ran\},
% &=\left\{\frac{|00\ran+|11\ran}{\sqrt{2}}|0\ran/|1\ran,|0\ran/|1\ran\frac{|00\ran+|11\ran}{\sqrt{2}},|010\ran,|101\ran\right\}\\
% &\quad\quad\quad \bigcup\left\{\frac{|000\ran-|011\ran-|110\ran}{\sqrt{3}},\frac{|111\ran-|001\ran-|100\ran}{\sqrt{3}}\right\}.
% \end{aligned}
\end{equation}
which satisfies $|\Omega(3)|=8=2^3$, and we verify that the basis matrix is full rank. Now, we assume that $\Omega(L)$ is linearly-independent for $L\leq x$. When we introduce a new qubit, we can classify $\Omega(x+1)$ into subsets depending on the entanglement structure of the extended system. If the new qubit is still \textit{disentangled} from the original chain, it forms a subset 
\begin{equation}
\Omega_0(x+1)=\Omega(x)\setminus\{|\psi\ran\in\Omega(x):|\psi\ran=|\dots\tikz[]{
\draw[dash pattern=on 3pt off 1.5pt,thick,NavyBlue] (0.06,0.035)--(0.54,0.035);
\draw[dash pattern=on 3pt off 1.5pt,thick,BrickRed] (0.06,-0.035)--(0.54,-0.035);
\fill[black] (0,0) circle (2pt);
\fill[black] (0.5,0) circle (2pt);
}\ran, |\psi\ran=|\dots\tikz[]{\fill[NavyBlue] (0,0) circle (2pt);}/\tikz[]{
\fill[BrickRed] (0,0) circle (2pt);}\ran\}\otimes|\tikz[]{
\fill[NavyBlue] (0,0) circle (2pt);}\ran/|\tikz[]{
\fill[BrickRed] (0,0) circle (2pt);}\ran. 
\end{equation}
\end{widetext}
The states ending with an entangled frozen state or a ``dot'' identical with the new qubit are excluded from $\Omega(x)$, as states of the form $|\dots\tikz[]{
\draw[dash pattern=on 3pt off 1.5pt,thick,NavyBlue] (0.06,0.035)--(0.54,0.035);
\draw[dash pattern=on 3pt off 1.5pt,thick,BrickRed] (0.06,-0.035)--(0.54,-0.035);
\fill[black] (0,0) circle (2pt);
\fill[black] (0.5,0) circle (2pt);
}\ran\otimes|\tikz[]{
\fill[NavyBlue] (0,0) circle (2pt);}\ran/|\tikz[]{
\fill[BrickRed] (0,0) circle (2pt);}\ran$ and $|\dots\tikz[]{
\fill[NavyBlue] (0,0) circle (2pt);
\fill[NavyBlue] (0.5,0) circle (2pt);}\ran$ are no longer frozen. It is easy to see that $\Omega_0$ is linearly-independent. Next, we introduce another linearly-independent subset if the new qubit forms a dimer with site $x$,
\begin{equation}
    \Omega_1(x+1) = \Omega(x-1)\otimes|\tikz[]{\fill[black] (0,0) circle (2pt);
\fill[black] (0.5,0) circle (2pt);
\draw[thick,black] (0,0)--(0.5,0);}\ran.
\end{equation}
Since the states in $\Omega_1$ are maximally-entangled across the bond $(x,x+1)$, they do not overlap with states in $\Omega_0$. Therefore, $\Omega_0\cup\Omega_1$ is linearly-independent. In addition, the new qubit can extend the entangled frozen regime of length $n-1$ to $n$ at the end of the chain, forming a series of subsets
\begin{equation}
    \Omega_n(x+1) = \{|\psi\ran\in\Omega(x-n+1): |\dots\tikz[]{\fill[black] (0,0) circle (2pt);
\fill[black] (0.5,0) circle (2pt);
\draw[thick,black] (0,0)--(0.5,0);}\ran\}\otimes|\tikz[]{
\draw[dash pattern=on 3pt off 1.5pt,thick,NavyBlue] (0.06,0.035)--(0.54,0.035);
\draw[dash pattern=on 3pt off 1.5pt,thick,BrickRed] (0.06,-0.035)--(0.54,-0.035);
\fill[black] (0,0) circle (2pt);
\fill[black] (0.5,0) circle (2pt);}
\dots\tikz[]{
\draw[dash pattern=on 3pt off 1.5pt,thick,NavyBlue] (0.06,0.035)--(0.54,0.035);
\draw[dash pattern=on 3pt off 1.5pt,thick,BrickRed] (0.06,-0.035)--(0.54,-0.035);
\fill[black] (0,0) circle (2pt);
\fill[black] (0.5,0) circle (2pt);
}\ran_n
\end{equation}
We require the configuration before the entangled frozen regime to be a dimer for the same reason as above. If a state $|\psi\ran\in\Omega_2$ can be written as a linear combination of the states in $\Omega_0\cup\Omega_1$, the contribution from $\Omega_0$ must involve states of the form $|\dots\tikz[]{\fill[black] (0,0) circle (2pt);
\fill[black] (0.5,0) circle (2pt);
\draw[thick,black] (0,0)--(0.5,0);}\ran\otimes|\tikz[]{
\fill[NavyBlue] (0,0) circle (2pt);
\fill[BrickRed] (0.5,0) circle (2pt);}\ran/|\tikz[]{
\fill[NavyBlue] (0.5,0) circle (2pt);
\fill[BrickRed] (0,0) circle (2pt);}\ran$, since other possible states in $\Omega_0$ such as $|\dots\tikz[]{\fill[black] (0,0) circle (2pt);
\fill[black] (0.5,0) circle (2pt);
\draw[thick,black] (0,0)--(0.5,0);
\fill[NavyBlue] (1,0) circle (2pt);
\fill[BrickRed] (1.5,0) circle (2pt);
\fill[NavyBlue] (2,0) circle (2pt);}\ran$ and $|\dots\tikz[]{
\fill[NavyBlue] (0.5,0) circle (2pt);
\fill[BrickRed] (0,0) circle (2pt);
\fill[BrickRed] (1,0) circle (2pt);
\fill[NavyBlue] (1.5,0) circle (2pt);}\ran$ do not recover the entanglement structure of $|\psi\ran$ with a dimer on $(x-2,x-1)$. Similarly, the contribution from $\Omega_1$ must involve states of the form $|\dots\tikz[]{\fill[black] (0,0) circle (2pt);
\fill[black] (0.5,0) circle (2pt);
\draw[thick,black] (0,0)--(0.5,0);}\ran\otimes |\tikz[]{\fill[black] (0,0) circle (2pt);
\fill[black] (0.5,0) circle (2pt);
\draw[thick,black] (0,0)--(0.5,0);}\ran$. However, these states are all orthogonal to $|\psi\ran$, so we conclude that $\Omega_0\cup\Omega_1\cup\Omega_2$ is linearly-independent. Following similar reasoning, $\Omega_n$ are then introduced into the basis one by one until $n=x+1$ (where the whole state is an entangled frozen state). Finally, we can show that the basis 
\begin{equation}
    \Omega(x+1)=\bigcup_{n=0}^{x+1}\Omega_n(x+1)
\end{equation}
is linearly-independent given that $\Omega(x)$ is linearly-independent. Therefore, $\Omega(L)$ is linearly-independent for generic $L$. 

We use the same induction method to prove the completeness of the basis for $N=2$. If a basis state of length $L$ has $N_d$ dimers, then there are $N_d+1$ gaps where we can insert frozen regimes of length $d_i\in\mathbb{N}$, with $\sum_{i=1}^{N_d} d_i=L-2N_d$. For each frozen regime of length $d_i$, there are 2 frozen product states, and $d_i-1$ entangled frozen states as mentioned before, giving a total of $d_i+1$ frozen states. The number of configurations within the basis is thus
\begin{equation}
    |\Omega(L)|=\sum_{N_d=0}^{\lfloor L/2\rfloor}\sum_{\{d_i:\sum_{i}d_i=L-2N_d\}} \prod_{i=1}^{N_d+1}(d_i+1).
\end{equation}
Let us assume that $|\Omega(L)|=2^L$ for $L\leq x$, and we have already verified that this holds for $L\leq 3$. Without loss of generality, we assume that $x$ is even. When introducing a new qubit, we can again classify the configurations into two cases: 1. The new qubit extends the $(N_d+1)$-th frozen regime by one site. 2. The new qubit forms a dimer with the qubit on site $x$. This gives
\begin{widetext}
\begin{equation}
\begin{aligned}
    |\Omega(x+1)|&=\sum_{N_d=0}^{x/2}\sum_{\{d_i\}} \prod_{i=1}^{N_d}(d_i+1)(d_{N_d+1}+2)+\sum_{N_d=0}^{x/2-1}\sum_{\{d_i\}}\prod_{i=1}^{N_d+1}(d_i+1)\\
    &=\sum_{N_d=0}^{x/2}\sum_{\{d_i\}} \prod_{i=1}^{N_d+1}(d_i+1)+2\sum_{N_d=0}^{x/2-1}\sum_{\{d_i\}}\prod_{i=1}^{N_d+1}(d_i+1)\\
    &=|\Omega(x)|+2|\Omega(x-1)|=2^{x+1}.
\end{aligned}
\end{equation}
\end{widetext}
Therefore, $|\Omega(L)|=2^L$ for generic $L$, and it is a complete basis which spans the Hilbert space. \qed

\section{Dynamics of the tensor product of two entangled frozen segments of Temperley-Lieb model}\label{app: dynamics of two frozen segments}

In this Appendix, we analyze how a tensor product of two minimal entangled frozen segments,
\be |\psi^{s}_{l}\ran\otimes|\psi^{s'}_{l'}\ran, \ee
evolves under the Temperley-Lieb dynamics. Here, the irreducible strings are 
\bea s&=(a_1\dots a_{L_s}) \text{ with } a_i\in\{b,c\},\\
s'&=(a_1'\dots a'_{L_{s'}}) \text{ with } a_i'\in\{d,e\}. \eea
We assume that the two segments share at least one common color, i.e. 
\be \{b,c\}\cap\{d,e\}\neq\emptyset. \ee
The intuition is that local TL projectors act nontrivially only at the bond $(l,l+1)$, where identical colors meet. If the two frozen segments share a common color across the interface, successive applications of TL projectors convert matching pairs into dimers, reducing the state to a tensor product of dimers together with a frozen state labeled by irr$(ss')$. Thus, the evolution remains entirely within the Krylov sector determined by the concatenated irreducible string.

More explicitly, define the truncated string $s_{[i:j]}=(a_i\dots a_j)$. Since minimal entangled frozen states only involve two colors, we define $\overline{a_i}$ as the color different from $a_i$. Without loss of generality, assume that $l>l'$. We can decompose the state as
\begin{equation}\label{eq: decomposition}
    \begin{aligned}
        &|\psi^{s}_{l}\ran\otimes|\psi^{s'}_{l'}\ran \\
        =&(|\psi^{s_{[1:L_s-1]}}_{l-1}\ran\otimes|a_{L_s}\ran-|\psi^{s\overline{a_{L_s}}}_{l-1}\ran\otimes|\overline{a_{L_s}}\ran)\\
        &\otimes(|a'_1\ran\otimes|\psi^{s'_{[2:L_s']}}_{l'-1}\ran-|\overline{a'_1}\ran\otimes|\psi^{\overline{a'_1}s'}_{l'-1}\ran)
    \end{aligned}
\end{equation}
To see why this holds, recall that $|\psi^{s}_{l}\ran$ is the equal-weight superposition of states that involve only colors $\{b,c\}$ dynamically connected to the seed \be |a_1\dots a_{L_s}\overbrace{b\dots b}^\text{even}\ran. \ee
In particular, the superposition naturally separates into contributions in which the rightmost site carries
$a_{L_s}$ (e.g., $|a_1\dots a_{L_s-1}b\dots ba_{L_s}\ran$), producing 
\be |\psi^{s_{[1:L_s-1]}}_{l-1}\ran\otimes|a_{L_s}\ran,\ee 
and contributions in which it carries the complementary color $\overline{a_{L_s}}$ at the end (e.g., $-|a_1\dots a_{L_s}b\dots b\overline{a_{L_s}} \overline{a_{L_s}}\ran$), giving
\be -|\psi^{s\overline{a_{L_s}}}_{l-1}\ran\otimes|\overline{a_{L_s}}\ran.\ee
Similar decomposition applies to the second state.

We first discuss the case when $\{b,c\}=\{d,e\}$, i.e., the two frozen segments involve the same pair of colors. Suppose further that $a_1'=a_{L_s}$, then $\overline{a_1'}=\overline{a_{L_s}}$. Since adjacent entries in an irreducible string must differ, we then have $a_2'=a_{L_s-1}$. Continuing this argument inductively, we find that $a_j'=a_{L_s-j+1}$, i.e., the second irreducible string is the reverse of the tail of the first. Therefore, in the concatenated sequence $ss'$, adjacent entries cancel pairwise from the interface inward, yielding
\be \text{irr}(ss')=
\begin{cases}
    (a_1\dots a_{L_s-L_{s'}+1}), \text{ if $L_s>L_{s'}$}\\
    (a'_{L_{s'}-L_s+1}\dots a'_{L_{s'}}), \text{ else}.
\end{cases} \ee
The projector $\hat{P}_{l,l+1}$ then maps Eq.~\eqref{eq: decomposition} to
\bea
(|\psi^{s_{[1:L_s-1]}}_{l-1}\ran\otimes|\psi^{s'_{[2:L_s']}}_{l'-1}\ran+|\psi^{s\overline{a_{L_s}}}_{l-1}\ran\otimes|\psi^{\overline{a'_1}s'}_{l'-1}\ran)\otimes|\tikz[]{\fill[black] (0,0) circle (2pt);
\fill[black] (0.5,0) circle (2pt);
\draw[thick,black] (0,0)--(0.5,0);}\ran.
\eea
Each application of the projector removes one matching pair at the interface, generating one dimer. If $l$ and $l'$ are large enough, the decomposition at time $t$ will give rise to terms ranging from 
\be |\psi^{s_{[1,L_s-t]}}_{l-t}\ran\otimes|\psi^{s'_{[t+1:L_s']}}_{l'-t}\ran\ee to \be|\psi^{s\overline{a_{L_s}\dots a_{L_s-t+1}}}_{l-t}\ran\otimes|\psi^{\overline{a'_t\dots a'_1}s'}_{l'-t}\ran.\ee
This process can be continued until either 
\begin{enumerate}
    \item  one of the irreducible strings becomes empty, or
    \item one of the segments becomes a frozen product state.
\end{enumerate}
Without loss of generality, assume $L_s>L_{s'}$ and $l-L_s>l'-L_{s'}$, so that the second segment is progressively ``absorbed'' into the first. 

In the first case, suppose  at time $t=T$ the second segment becomes a frozen state with an empty irreducible string. It is then a superposition of product states containing an even number of $b$ and $c$ spins, and its length $l'(T)$ must be even. In this situation, the first segment evolves into an entangled frozen state with the same irreducible string as $\text{irr}(ss')$ after ``merging'' with the second by constantly applying projection operators at the interface.

The second case corresponds to the second segment becoming a frozen product state at time $T$, e.g., \be|\psi^{\overline{a'_T\dots a'_1}s'}_{l'-T}\ran=|\overline{a'_T\dots a'_1}a'_1\dots a'_{L_{s'}}\ran, \text{ with }T=\frac{l'-L_{s'}}{2}.
\ee 
Assume $|\text{irr}(ss')|< l-l'$. If we continue decomposing the first segment $|\psi^{s\overline{a_{L_s}\dots a_{L_s-T+1}}}_{l-T}\ran$, since the second segment is now a fixed product state, all other superposition components are annihilated by the projector, and the only term that contributes nontrivially is
\be|\psi^{\text{irr}(ss')}_{l-l'}\ran\otimes|a'_{L_{s'}}\dots a'_1\overline{a_{L_s}\dots a_{L_s-T+1}}\ran,\ee
successive applications of the projector at the interface maps 
\be |a'_{L_{s'}}\dots a'_1\overline{a_{L_s}\dots a_{L_s-T+1}}\ran\otimes|\overline{a'_T\dots a'_1}a'_1\dots a'_{L_{s'}}\ran \ee
to $T+L_{s'}$ dimers since we assumed $a_j'=a_{L_s-j+1}$ in the beginning and the two strings cancel pairwise from the interface inward. Together with $T$ dimers generated at time $t=T$, there are $2T+L_{s'}=l'$ dimers in the final state. Both scenarios lead to the state
\be |\psi^{\text{irr}(ss')}_{l-l'}\ran\otimes(|\tikz[]{
\draw[thick,Black] (0,0)--(0.5,0);
\fill[black] (0,0) circle (2pt);
\fill[black] (0.5,0) circle (2pt);}\ran)^{\otimes l'}, \text{ if $|\text{irr}(ss')|< l-l'$}. \ee
If instead $|\text{irr}(ss')|\geq l-l'$, the state is mapped to
\be
 |\text{irr}(ss')\ran\otimes(|\tikz[]{
\draw[thick,Black] (0,0)--(0.5,0);
\fill[black] (0,0) circle (2pt);
\fill[black] (0.5,0) circle (2pt);}\ran)^{\otimes \frac{l+l'-|\text{irr}(ss')|}{2}}.
\ee

Alternatively, if $a_1'=\overline{a_{L_s}}$ and hence $\overline{a'_1}=a_{L_s}$, we have $a_j\neq a_j'$ and hence $\text{irr}(ss')=ss'$. The projector $\hat{P}_{l,l+1}$ maps Eq.~\eqref{eq: decomposition} to
\bea
-(|\psi^{s_{[1:L_s-1]}}_{l-1}\ran\otimes|\psi^{a_{L_s}s'}_{l'-1}\ran+|\psi^{sa'_1}_{l-1}\ran\otimes|\psi^{s'_{[2:L_s']}}_{l'-1}\ran)\otimes|\tikz[]{\fill[black] (0,0) circle (2pt);
\fill[black] (0.5,0) circle (2pt);
\draw[thick,black] (0,0)--(0.5,0);}\ran
\eea
and the same reasoning leads to the identical final structure.

If $\{b,c\}\neq\{d,e\}$, say $b=d$ but $e\notin\{b,c\}$, then either $s'=(bebe\dots)$ or $s'=(ebeb\dots)$. For the former, the projector $\hat{P}_{l,l+1}$ only acts nontrivially on the term $|\psi^{s}_{l}\ran\otimes |b\ran\otimes|\psi^{(ebe\dots)}_{l'-1}\ran$ and maps Eq.~\eqref{eq: decomposition} to 
\be |\psi^{\text{irr}(sa_1')}_{l-1}\ran\otimes|\psi^{s'_{[2:L_s']}}_{l'-1}\ran\otimes|\tikz[]{\fill[black] (0,0) circle (2pt);
\fill[black] (0.5,0) circle (2pt);
\draw[thick,black] (0,0)--(0.5,0);}\ran \ee
Under the application of another projection operator on the bond between the two frozen segments, the state is mapped to
\be\label{eq: ss'} |\psi^{s}_{l-2}\ran\otimes|\psi^{s'}_{l'-2}\ran\otimes(|\tikz[]{\fill[black] (0,0) circle (2pt);
\fill[black] (0.5,0) circle (2pt);
\draw[thick,black] (0,0)--(0.5,0);}\ran)^{\otimes 2} \ee
We can continue this process until either one of the segments becomes a frozen product state. If $l-L_s>l'-L_{s'}$, then the second segment first becomes a product state, i.e. $|\psi^{s'}_{L_{s'}}\ran=|a'_1\dots a'_{L_{s'}}\ran$. Since $a'_1\in\{b,c\}$, we can apply the projector again and obtain
\be |\psi^{\text{irr}(sa_1')}_{l-l'+L_{s'}-1}\ran\otimes|a'_2\dots a'_{L_{s'}}\ran\otimes(|\tikz[]{\fill[black] (0,0) circle (2pt);
\fill[black] (0.5,0) circle (2pt);
\draw[thick,black] (0,0)--(0.5,0);}\ran)^{\otimes l'-L_{s'}+1}. \ee
If instead $l-L_s<l'-L_{s'}$, then the first segment first becomes a product state, i.e. $|\psi^{s}_{L_s}\ran=|a_1\dots a_{L_s}\ran$. Similarly, applying the projector between the two segments gives
\be |a_1\dots a_{L_s-1}\ran\otimes|\psi^{\text{irr}(a_{L_s}s')}_{l'-l+L_s-1}\ran\otimes(|\tikz[]{\fill[black] (0,0) circle (2pt);
\fill[black] (0.5,0) circle (2pt);
\draw[thick,black] (0,0)--(0.5,0);}\ran)^{\otimes l-L_{s}+1}. \ee
If $s'=(ebeb\dots)$, the projector only acts nontrivially on the term $-|\psi^{s}_{l}\ran\otimes |b\ran\otimes|\psi^{(bebe\dots)}_{l'-1}\ran$ and maps Eq.~\eqref{eq: decomposition} to 
\be |\psi^{\text{irr}(s\overline{a_1'})}_{l-1}\ran\otimes|\psi^{\overline{a_1'}s'}_{l'-1}\ran\otimes|\tikz[]{\fill[black] (0,0) circle (2pt);
\fill[black] (0.5,0) circle (2pt);
\draw[thick,black] (0,0)--(0.5,0);}\ran \ee
The following projector maps to the same state Eq.~\eqref{eq: ss'}. This process can be repeated until one of the segments becomes a product state.  If $l-L_s>l'-L_{s'}$, then the second segment first becomes a product state, i.e. $|\psi^{\overline{a'_1}s'}_{L_{s'}+1}\ran=|\overline{a'_1}a'_1\dots a'_{L_{s'}}\ran$. The following application of the projection operator leads to
\be |\psi^{s}_{l-l'+L_{s'}}\ran\otimes|s'\ran\otimes(|\tikz[]{
\draw[thick,Black] (0,0)--(0.5,0);
\fill[black] (0,0) circle (2pt);
\fill[black] (0.5,0) circle (2pt);}\ran)^{\otimes l'-L_{s'}}. \ee
Similarly, if $l-L_s<l'-L_{s'}$, the state is connected to
\be |s\ran\otimes|\psi^{s'}_{l'-l+L_s}\ran\otimes(|\tikz[]{
\draw[thick,Black] (0,0)--(0.5,0);
\fill[black] (0,0) circle (2pt);
\fill[black] (0.5,0) circle (2pt);}\ran)^{\otimes l-L_{s}}. \ee

\bibliographystyle{apsrev4-2}
\bibliography{reference}
\end{document}